\documentclass[twocolumn,prb,superscriptaddress]{revtex4}
\usepackage{graphicx}
\usepackage{color}\usepackage{enumerate}
\usepackage{amsmath}\usepackage{amssymb}\usepackage{stmaryrd}
\usepackage{esint}
\usepackage{multirow}
\usepackage{float}
\usepackage{longtable,booktabs}
\usepackage{subfigure}
\usepackage{dsfont}
\usepackage{amssymb}
\usepackage{comment}
\usepackage[pdftex,colorlinks=true,linkcolor=blue,citecolor=blue]{hyperref}

\usepackage{mathtools}
\DeclarePairedDelimiter\bra{\langle}{\rvert}
\DeclarePairedDelimiter\ket{\lvert}{\rangle}
\DeclarePairedDelimiterX\braket[2]{\langle}{\rangle}{#1 \delimsize\vert #2}

\usepackage{tikz}
\def\quadplus{
    \tikz[baseline=0.4ex]{
        \draw (0,0) -- (0,2ex) -- (2ex,2ex) -- (2ex,0) -- (0,0);
        \fill (0,0) circle (2pt);
        \fill (2ex,2ex) circle (2pt);
        \draw[fill=white] (0,2ex) circle (2pt);
        \draw[fill=white] (2ex,0) circle (2pt);
    }
}
\def\quadminus{
    \tikz[baseline=0.4ex]{
        \draw (0,0) -- (0,2ex) -- (2ex,2ex) -- (2ex,0) -- (0,0);
        \fill (0,2ex) circle (2pt);
        \fill (2ex,0) circle (2pt);
        \draw[fill=white] (2ex,2ex) circle (2pt);
        \draw[fill=white] (0,0) circle (2pt);
    }
}

\begin{document}
\title{Higher-form Gauge Symmetries in Multipole Topological Phases}

\author{Oleg Dubinkin\footnote{These authors contributed equally to the development of this work.}}
\affiliation{Department of Physics and Institute for Condensed Matter Theory, University of Illinois at Urbana-Champaign, IL 61801, USA}
\author{Alex Rasmussen\footnotemark[1]}
\affiliation{Department of Physics, The Ohio State University, Columbus, OH 43210, USA}
\author{Taylor L. Hughes}
\affiliation{Department of Physics and Institute for Condensed Matter Theory, University of Illinois at Urbana-Champaign, IL 61801, USA}
\begin{abstract}
In this article we study field-theoretical aspects of multipolar topological insulators. Previous research has shown that such systems naturally couple to higher-rank tensor gauge fields that arise as a result of gauging dipole or subsystem $U(1)$ symmetries. Here we propose a complementary framework using electric higher-form symmetries. We utilize the fact that gauging 1-form electric symmetries results in a 2-form gauge field which couples naturally to extended line-like objects: Wilson lines. In our context the Wilson lines are electric flux lines associated to the electric polarization of the system. This allows us to define a generalized 2-form Peierls' substitution for dipoles that shows that the off-diagonal components of a rank-2 tensor gauge field $A_{ij}$ can arise as a lattice Peierls factor generated by the background antisymmetric 2-form gauge field. This framework has immediate applications: (i) it allows us to construct a manifestly topological quadrupolar response action given by a Dixmier-Douady invariant -- a generalization of a Chern number for 2-form gauge fields -- which makes plain the quantization of the quadrupole moment in the presence of certain crystal symmetries; (ii) it allows for a clearer interpretation of the rank-2 Berry phase calculation of the quadrupole moment; (iii) it allows for a proof of a generic Lieb-Schultz-Mattis theorem for dipole-conserving systems.
\end{abstract}

\maketitle

\section{Introduction}

Symmetry protected topological (SPT) phases of matter have been of fundamental interest since the discovery of topological insulators.\cite{kane2005,hasan2010,hasan2011,maciejko2011,senthil2015,chen2013,wang2013,kapustin2014,bi2015}
For systems protected by only internal symmetries, extensive analyses have yielded a nearly complete understanding of the topological classification and the associated physical phenomena. 
One of the fundamental features of such phases is the existence of gapless modes at the boundary, protected by this internal symmetry.
In recent years, there was a surge of interest towards a class of systems protected by spacetime symmetries realizing what was named higher-order symmetry protected topological phases (HOSPT).\cite{benalcazar2017quantized,benalcazar2017electric,hoti2018,Trifunovic2019,song2017a,song2017b,wang2018,Langbehn2017,teo2013,Isobe15,benalcazar2014,benalcazar2018,benalcazar19,you2018,varjas19,dmmh2019dipole} 
In a non-trivial phase, such models may host protected gapless modes only at subdimensional edges of the lattice, e.g., corners and hinges, while the rest of the boundary stays gapped.
General classification schemes and proposals for further generalization of such phases quickly followed,\cite{rasmussen2018a,thorngren18,Calugaru19,else2019} which included phases protected by a combination of spacetime and internal symmetry.

A subclass of HOTIs are the electric multipole insulators. The first example is the so-called quadrupole topological insulator \cite{benalcazar2017quantized,benalcazar2017electric} (QTI), which is a free-fermion system that exhibits corner charges due to the presence of the quantized quadrupole moment in the bulk. Importantly, for the bulk quadrupole moment to be a well-defined quantity, one must require that both the total electric charge and the overall charge polarization of the system vanish at all times. In the insulating phases of the QTI, the polarization of the ground state is quantized/fixed by spatial symmetries. However, in more recent developments the concept of a quadrupole HOTI has been extended to systems that exhibit both microscopic charge and dipole conservation.\cite{ybh2019,dmmh2019dipole} In such systems the charge and polarization of the ground state and all excited states are fixed quantum numbers, and the quadrupole moment is well-defined if they both vanish. While much is known about the quadrupole HOTIs, one element for which there are still open questions is the connection between the topological classification and the observable, quantized quadrupole moment response. There have been several articles that address key aspects of this issue\cite{benalcazar2017quantized,benalcazar2017electric,ybh2019,dmmh2019dipole} that we will review below, and in this article we will develop a new approach that helps resolve some subtleties that arose in the previous works, and finds some other immediate applications.

A useful framework for studying dipole-conserving models and topological quadrupolar response is a field-theoretical point of view that was developed to describe recently uncovered fracton models. These models contain symmetric tensor gauge fields, dubbed higher rank gauge fields. 
In particular, so-called scalar-charge theories couple to a rank-2 $U(1)$ gauge field $A_{ij},$ and are equipped with the following rank-2 Gauss' law constraint:   
\begin{equation}
    \partial_i \partial_j E_{ij} = 0,
\end{equation}\noindent where $E_{ij}=\partial_i \partial_j A_0-\partial_t A_{ij}$ is the rank-2 electric field. The rank-2 Gauss' law leads to both the conservation of the total charge and the total dipole moment of the system to which the gauge field couples. 
Furthermore, theories where matter couples to only the off-diagonal components of the rank-2 gauge field, e.g., $A_{xy}$ support a new class of global symmetries that give rise to $U(1)$ charge conservation on spatial subsystems.\cite{pretko2018gauge} 
On the lattice, such off-diagonal rank-2 tensor fields naturally couple to dynamical ring-exchange processes,\cite{ybh2019,dmmh2019dipole} analogous to how single-particle tunneling terms naturally couple to the rank-1 electromagnetic vector potential via Peierls' substitution. Models built from ring-exchange dynamics that have microscopic dipole conservation have been shown to exhibit HOTI phases including the aforementioned quantized quadrupole phase,\cite{ybh2019,dmmh2019dipole}
Furthermore, in Ref. \onlinecite{dmmh2019dipole}, these ring-exchange models were coupled to rank-2 tensor gauge fields and used to define a rank-2 Berry phase that  successfully distinguishes between the topological and trivial phases that differ by their bulk quadrupole moments. This is in exact analogy to how the Berry phase defined with the help of the regular electromagnetic $U(1)$ gauge field can determine the electric polarization.\cite{kingsmith93,ortiz1994}

\begin{figure}
    \centering
    \includegraphics[width=0.5\textwidth]{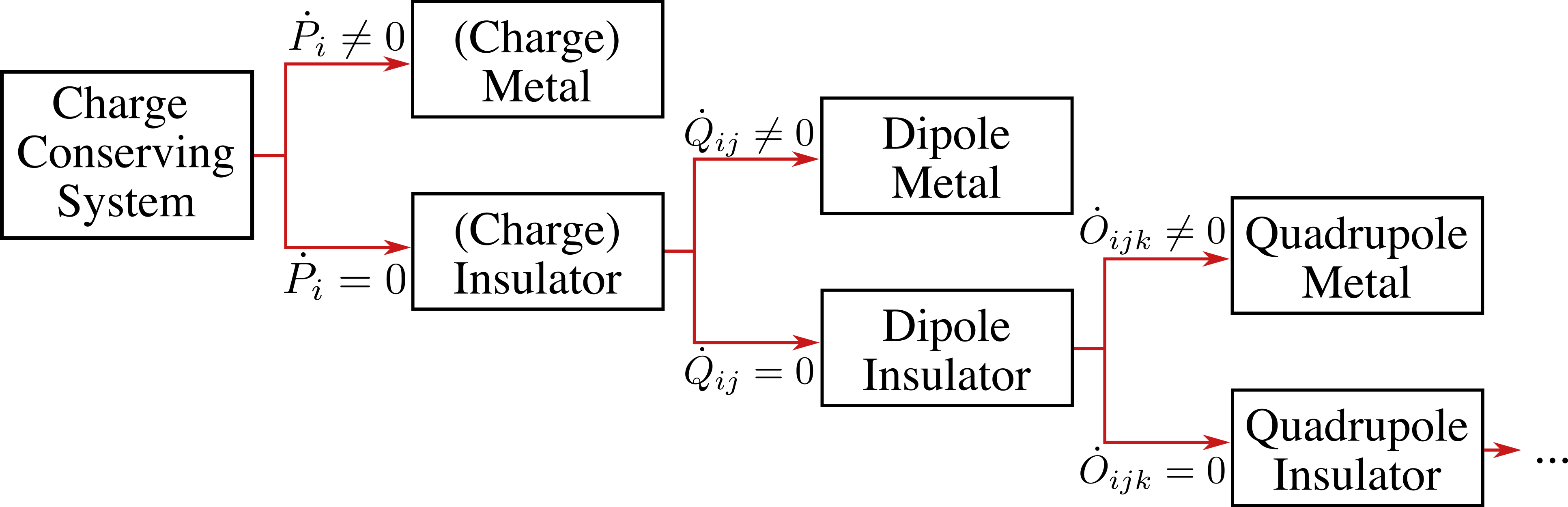}
    \caption{A hierarchical refinement of the notion of `insulator.'}
    \label{fig:insulator_tree}
\end{figure}

Let us now consider a complementary framework to approach such dipole-conserving models.
Recently, Ref. \onlinecite{dmmh2019dipole} proposed a refinement of the notion of an insulator. 
In the absence of electrical currents, $\vec{J}=0$, the electrical polarization, i.e., the overall $1$-st multipole moment of the system, is conserved 
\begin{equation}
    \frac{dP_i}{dt}=J_i=0.
\end{equation}
With this in mind, one then may ask whether the quadrupole moment $Q_{ij}$ of the system, i.e., the overall $2$-nd multipole moment, is also conserved in the physical system at hand. In the absence of electrically charged currents, the quadrupole moment of the system can in principle still be altered by currents carrying electric dipole moments. Heuristically, we can then say that a charge insulator which satisfies $\dot{Q}_{ij}\neq 0$ will support `dipole currents,' and we can call such systems \emph{dipole metals}. In the opposite case, when $\dot{Q}_{ij}=J_{ij}=0$, where $J_{ij}$ is the dipole current, the system is a \emph{dipole insulator}. One can then ask whether the system supports currents $J_{ijk}$ that carry quadrupole moments that would change  the octupole moment $O_{ijk},$ and so on. This chain of reasoning is illustrated in Fig. \ref{fig:insulator_tree} from which one can see that an $n$-th order multipole insulator must conserve the $0$-th through $(n+1)$-th multipole moments with the $0$-th multipole moment being the total electric (monopole) charge.
In the present article, we use this foundation to propose a framework that treats each conserved $n$-th multipole moment as a conserved charge corresponding to an $n$-form symmetry in the system. We will focus our attention on systems conserving overall dipole moment, which we will treat as a conserved charge of a global electric 1-form symmetry that exists in the absence of electrical charge currents. In this approach, the electric field associated to the 1-form symmetry is essentially the internal field produced by the polarized material, which provides a natural connection to dipole conservation. Such higher-form symmetries have been of particular interest in the high-energy literature.\cite{gukov2008,Gaiotto2015,lake2018,rasmussen2018c,hawking2016} Quite recently, these concepts were applied in condensed matter systems to study fracton phenomena,\cite{seiberg2019,seiberg2020a,seiberg2020b,seiberg2020c,radicevic2019,jwang1,jwang2} which are in turn related\cite{ybh2019} to the concept of higher order multipole topological insulators. 

 A central feature in our framework is the necessary introduction of an \textit{antisymmetric} 2-form tensor field $\mathcal{B}.$ This is necessitated by the gauging of the global 1-form symmetry that corresponds to the conservation of the total polarization of the system.  Our formalism bridges the gap between lattice models that exhibit dipole conservation due to the presence of subsystem $U(1)$ symmetries, and field theories with global 1-form symmetries,  both of which act on co-dimension 1 manifolds. The field $\mathcal{B}$ naturally couples to one-dimensional extended objects, i.e., the electric lines associated to the dipole moment, and imparts a new type of Peierls' phase to the motion of dipoles. A key result is a new interpretation of the lattice field $A_{xy}$ as a phase that arises from the motion of dipoles in the background 2-form field $\mathcal{B}$. The relationship between $A_{xy}$ and $\mathcal{B}$ is captured by the generalized Peierls' substitution:
\begin{equation}
    A_{xy}=\frac{e}{\hbar}\iint_{p_{xy}} \mathcal{B},
    \label{eqn:gen_p0}
\end{equation}
where $p_{xy}$ denotes an elementary plaquette in the spatial lattice plane.
As we will see below, one interesting consequence of this approach is that it resolves a subtlety from Ref. \onlinecite{dmmh2019dipole} by making apparent the topological nature of the quadrupolar rank-2 Berry phase mentioned earlier. Indeed, the rank-2 Berry phase\cite{dmmh2019dipole} that is picked up by the ground state during the adiabatic insertion of one unit of rank-2 flux through the system, can be reinterpreted as a process that changes $\mathcal{B}$ by a global gauge transformation. Additionally, the quadrupolar $\theta$-term response of Ref. \onlinecite{ybh2019} can be reinterpreted as the topological response
\begin{eqnarray}
S_Q&=&-\frac{e\theta}{2\pi}\int d\mathcal{B},
\end{eqnarray} which is proportional to the Dixmier-Douady number of the $\mathcal{B}$ field configuration,\cite{Murray96,Nakahara2003,Baez2011} and provides a simple route to proving the quantization of the quadrupolarization $q_{xy}=e\theta/2\pi$.

Our paper is structured as follows: we start with Section \ref{sec:modelreview} where  motivate a topological formulation of the rank-2 Berry phase by considering the parallels between the  Berry phase of the SSH model and the rank-2 Berry phase which was proposed for a ring-exchange quadrupole model. Then in Section \ref{sec:review_gauging} we review higher form symmetries and formalize polarization as a conserved charge of a 1-form symmetry. The main formal results are presented in Section \ref{sec:peierls} where we develop a generalized Peierls' substitution for dipole-conserving models. In Sections \ref{sec:response} and \ref{sec:lsm} we explore two applications of our formalism, the topological quadrupole response, and a  generalization of the Lieb-Schultz-Mattis theorem to systems that conserve dipole moments respectively.

\section{Berry phases and Topological Response in Dipole and Quadrupole Models}\label{sec:modelreview}
\label{sec:review_Berry}
One of the main results of this paper will be the development of a topological response term for the quadrupolarization in dipole-conserving systems. 
To set the context for the rest of this paper, it is worth reviewing a simple dipole-conserving lattice model which was shown\cite{ybh2019} to exhibit such a quantized response. However, before getting to this model we will begin by reviewing the features of a well-known charge-conserving model of a polarized 1D chain with a topological response term for charge polarization. Our presentation will exploit the analogies between the latter model, which has a dipole polarization, and the former model, which has a quadrupole polarization.

\subsection{Dipole model}
Let us start with a well-known Su-Schrieffer-Heeger (SSH) model that describes a one-dimensional chain of non-interacting spinless fermions with the following Hamiltonian:
\begin{equation}
    \mathcal{H}_{SSH}=\sum_{i=1}^N (t c^\dagger_{i,A} c_{i,B}+\lambda c^\dagger_{i,B}c_{i+1,A} + h.c.),
    \label{eqn:ssh_hamiltonian}
\end{equation}
where $A,B$ are two degrees of freedom in the unit cell, we assume periodic boundary conditions $N+1\equiv 1,$ and the model has inversion symmetry $x\to -x$.
When the magnitudes of the intra-cell and inter-cell couplings $t$ and $\lambda$ are not equal, this model is gapped and becomes an insulator at half-filling. 

The two insulating phases of this model, defined by the coupling ratios $|\lambda/t|>1,$ and $|\lambda/t|<1$, respectively can be distinguished by a global property -- a Berry phase computed around the non-contractible loop of the Brillouin zone.\cite{zak1989,hughes2011} Let us illustrate this result. Since we are working with a free-fermion model with translational invariance, we can write its ground state as a Slater determinant of Bloch orbitals, which are ambiguous up to the gauge transformation:
\begin{equation}
    \ket{u(k)}\to \text{e}^{i\phi(k)}\ket{u(k)},
    \label{eqn:gauge_transf}
\end{equation}
where the phase satisfies periodic boundary conditions: $\phi(\pi)-\phi(-\pi)=2\pi n$. This gauge structure allows for the definition of the Berry connection: 
\begin{equation}
    \mathfrak{a}(k)=-i\bra{u(k)} \partial_k \ket{u(k)},
\end{equation}
which transforms under the gauge transformation (\ref{eqn:gauge_transf}) as: $\mathfrak{a}(k)\to \mathfrak{a}(k)+\partial_k \phi(k)$. We can then compute a gauge-invariant quantity: the Berry phase around a closed loop in a parameter space that coincides with the Brillouin zone:
\begin{equation}
    \gamma_{BZ}=\oint_{BZ}\mathfrak{a}(k)dk=\begin{cases} \pi\mod{2\pi}& \text{when}\ \left\vert\frac{\lambda}{t}\right\vert>1\\ 0\mod{2\pi}& \text{when}\ \left\vert\frac{\lambda}{t}\right\vert<1 \end{cases}.
    \label{eqn:berry_phase}
\end{equation} The quantization of $\gamma_{BZ}$ to take only the values $0,\pi$ is enforced by the inversion symmetry of Eq.  (\ref{eqn:ssh_hamiltonian}). A key physical difference between these two insulating phases can be observed by opening the boundary in a manner that preserves the unit cell structure and then observing that, in the phase having non-trivial Berry phase, the system hosts  fractional $\pm e/2$ charges as a consequence of the non-trivial bulk polarization. This is opposed to the phase with $|\lambda/t|<1$ where the ground state of the system is unpolarized.

Let us now provide a slightly different interpretation of the Berry phase. We will consider our system coupled to a background electromagnetic field. First, let us review the way a 1-form gauge field manifests itself in tight-binding lattice models that conserve global $U(1)$ charge. In the presence of an external electromagnetic field, lattice translation operators are modified by a Peierls' phase in order to be gauge invariant:
\begin{equation}
    c^\dagger_{\textbf{r}_2}c_{\textbf{r}_1}\to \text{e}^{iA_{\textbf{r}_1,\textbf{r}_2}}c^\dagger_{\textbf{r}_2}c_{\textbf{r}_1}\ \text{with}\ A_{\textbf{r}_1,\textbf{r}_2}=\frac{e}{\hbar}\int_{\textbf{r}_1}^{\textbf{r}_2} \mathcal{A},
    \label{eqn:Peierls_tb}
\end{equation}
where $\mathcal{A}$ is a 1-form electromagnetic field with components $\mathcal{A}_\mu$, and $A_{\textbf{r}_1,\textbf{r}_2}\equiv A_l$ is the phase associated with a particular link $l$ of the lattice having endpoints $\textbf{r}_1$ and $\textbf{r}_2$. 
Consider a one-dimensional, translationally-invariant tight-binding chain with periodic boundary conditions, and a gauge choice having a uniform vector potential $\mathcal{A}$ across the chain. In this case, the phase factor associated with every link of the chain is simply $\text{e}^{iA_l},$ where $A_l$ is the same for every link. 
Transforming our real-space Hamiltonian $\mathcal{H}(\mathcal{A})$ to momentum space gives the Bloch Hamiltonian $H(k,A_l)$. 
This provides a different interpretation for the phase factor $\text{e}^{iA_l}$ -- it now plays the role of a momentum shift as we simply have for the Bloch Hamiltonian:
\begin{equation}
    H(k,A_l)\equiv H(k+A_l,0).
\end{equation}

This equivalence allows us to understand the Berry phase computed across the Brillouin zone $\gamma_{BZ}$ as the Berry phase in the parameter space of the  uniform phases $A_l$, which are generated by the uniform external electromagnetic field $\mathcal{A}$. This approach can be used on systems beyond the free-fermion limit to calculate the charge polarization.\cite{ortiz1994,kingsmith93} To compute the more general Berry phase we need to sweep the phases $A_l$ across an effective Brillouin zone. Hence we let $A_l\to A_l+2\pi/L,$ or equivalently $\mathcal{A}_x\to \mathcal{A}_x+h/eL.$ Physically, this configuration change of the background gauge field is equivalent to one unit of magnetic flux threaded through the loop formed by our periodic chain.

We can now provide a physical interpretation of the Berry phase in Eq.  (\ref{eqn:berry_phase}): $\gamma_{BZ}$ is a phase difference acquired by the ground state wave-function under the adiabatic evolution during which we insert one unit of magnetic flux through the center of the periodic chain. During that evolution over a long time $T$, we can imagine $\mathcal{A}_x$ linearly interpolating between $\mathcal{A}_x=0$ and $\mathcal{A}_x= h/eL.$ Both field configurations with $\mathcal{A}_x=0$ and with $\mathcal{A}_x=h/eL$ are pure gauges that are related by a \emph{large} gauge transformation of the electromagnetic field. However, during the adiabatic process a constant electric field $E_x=-\partial_t \mathcal{A}_x(t)=-h/(eLT)$ is generated across the chain. 
For a static insulating phase, the total dipole moment $P_x$ is well-defined and couples to the non-vanishing electric field during this process. The result is a phase acquired during this evolution:
\begin{equation}
    \gamma_{BZ}=-\frac{1}{\hbar}\int_0^T dt \textbf{P}\cdot\textbf{E}=\frac{1}{\hbar}\int_0^T dt P_x\partial_t \mathcal{A}_x(t)=2\pi\frac{P_x}{eL}.\label{eq:bphasephys}
\end{equation} 
 Importantly, this analysis points to the symmetry-protected topological nature of the polarization as it is quantized in the presence of inversion symmetry. 
The polarization of the chain $p_x\equiv P_x/L$ is odd under inversion symmetry, but is defined only $\mod{e},$ and thus for an inversion symmetric polarized insulator we expect to find $\gamma_{BZ}=\pi\mod{2\pi},$ while for an unpolarized system we find $\gamma_{BZ}=0\mod{2\pi},$ as was shown to be the case for the SSH chain in Eq. (\ref{eqn:berry_phase}).

We can recast this polarization response in a field-theoretical language. 
For one-dimensional systems we can write the topological $\theta$-term,\cite{Goldstone81,Qi2011} which exactly represents the  polarization of the system $P$ coupled to electric field $E$:
\begin{equation}
    \label{eq:polar_resp0}
    S_{P}=\frac{e\theta}{2\pi}\int d\mathcal{A} \equiv \int dt dx\ p_xE_x,
\end{equation}
with the identification $p_x=\frac{e\theta}{2\pi}$. On a closed spacetime manifold where the polarization is uniform, this integral computes the Chern class of $\mathcal{A},$ and we have $\int d\mathcal{A}=2\pi n$ for some integer $n$.  This reveals the periodic nature of $\theta$: shifting it by $2\pi$ shifts the action by an integer multiple of $2\pi$, leaving the path integral invariant. Thus we have $\theta\approx\theta+2\pi$. This identification mirrors the ambiguity of the polarization for periodic systems. Furthermore, by requiring an invariance of the path integral with respect to spatial inversion symmetry, one quickly determines that the values of $\theta$ under this restriction can only be: $0$ or $\pi$ mod $2\pi$. 

Let us now see how many of these notions can be suitably carried over to models exhibiting dipole conservation and quadrupole moments.

\subsection{Quadrupole model}
\begin{figure}
    \includegraphics[width=0.4\textwidth]{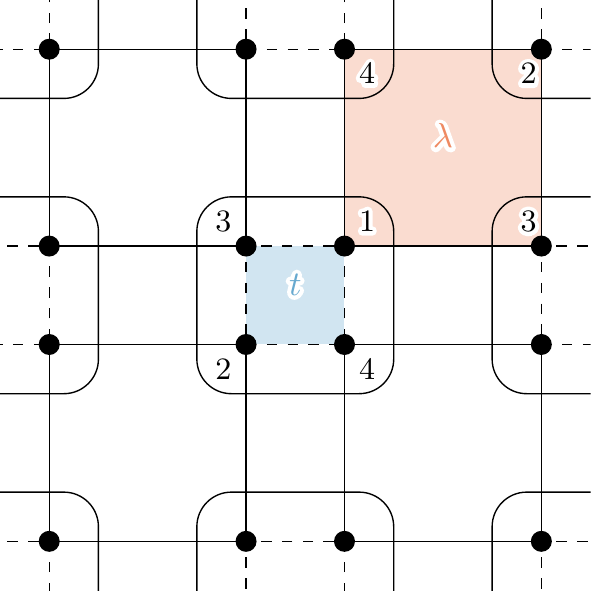}
    \caption{Lattice structure of the quadrupole ring-excahnge model.}
\label{fig:quadrupole_tb}
\end{figure}
In Refs. \onlinecite{ybh2019,dmmh2019dipole} it was shown that imposing a subsystem conservation of the $U(1)$ charge on top of the global $U(1)$ charge symmetry leads to a class of Hamiltonians that automatically conserve dipole moments and can support quadrupole moments. In particular, let us consider a two-dimensional model on a square lattice that independently conserves total charge along every row and every column. The basic allowed building blocks of the Hamiltonian are so-called ring-exchange terms that are associated with the plaquettes of the lattice:
\begin{equation}
    c^\dagger_{\textbf{r}+\hat{x}} c_{\textbf{r}} c^\dagger_{\textbf{r}+\hat{y}} c_{\textbf{r}+\hat{x}+\hat{y}}\equiv\left|\quadminus\right\rangle\left\langle\quadplus\right|,
    \label{eqn:dipole_hopping0}
\end{equation}
where white circles represent empty sites, and filled circles represent occupied sites. Taking these building blocks, consider now the \emph{ring-exchange quadrupole model} constructed purely from quartic interactions defined on a periodic $L_x\times L_y=N_x a\times N_y a$ lattice with four sites per unit cell as shown in Fig. \ref{fig:quadrupole_tb}:
\begin{equation}
    \begin{split}
    \mathcal{H}_{Q}=t\sum_{\textbf{r}}& (c^\dagger_{\textbf{r},2}c^{\phantom{\dagger}}_{\textbf{r},4}c^\dagger_{\textbf{r},1}c^{\phantom{\dagger}}_{\textbf{r},3}+h.c.)\\
    +&\lambda\sum_{\textbf{r}} (c^\dagger_{\textbf{r},1}c^{\phantom{\dagger}}_{\textbf{r}+\hat{x},3} c^{\dagger}_{\textbf{r}+\hat{x}+\hat{y},2}c^{\phantom{\dagger}}_{\textbf{r}+\hat{y},4}+h.c.).
    \end{split}
    \label{eqn:ring-exchange}
\end{equation}
Every term in this Hamiltonian conserves both the total particle number in the system, \emph{and} the total particle number along every column and every row of the lattice. Additionally, as a corollary, $\mathcal{H}_{Q}$ preserves the total polarization in both the $\hat{x}$ and $\hat{y}$ directions. It was argued\cite{dmmh2019dipole} that in the limits $|t|\ll|\lambda|$ or $|t|\gg|\lambda|$, the ground state of this model is gapped at half-filling, and the system is an insulator of both charges and dipoles. 
Furthermore, in both of these limits the fixed dipole moment is vanishing (at least up to a polarization quantum).

Since the system is both neutral and unpolarized its quadrupole moment is well-defined. Indeed it was shown in Refs. \onlinecite{ybh2019,dmmh2019dipole} that 
the quadrupolarization of the ring-exchange model is quantized in units of $q_{xy}=e/2.$
Moreover, it was also shown\cite{entanglement20} that in the zero-correlation length limit when $t=0$, the ground state of this model can be adiabatically connected, while preserving $C_4$ symmetry and hence vanishing polarization, to the ground state of the free-fermion quadrupole model.\cite{benalcazar2017quantized} Therefore, it is natural to expect that the quadrupolarization of the ring-exchange model is $q_{xy}=e/2$ when $\lambda$ dominates, and we expect it to be vanishing when $t$ dominates. 

To be more precise we can turn to two types of calculations to determine the quadrupole moment. We can consider the ground state expectation value of a many-body quadrupole operator,\cite{wheeler2018many,kang2018many} analogous to Resta's proposal for the charge polarization,\cite{resta1998} or we can use a Berry-phase calculation.\cite{dmmh2019dipole} For comparison to the previous section, let us consider the Berry phase approach. Our system is inherently interacting, so the idea of a Berry phase of Bloch functions over the periodic Brillouin zone is not applicable. Nevertheless, we can apply the more general procedure where we insert a uniform gauge field shift and determine the Berry phase of the ground state after the insertion of a flux quantum. However, the ring exchange model does not couple at all to a uniform electromagnetic vector potential $\mathcal{A}.$ To see this, one can resolve a ring-exchange term as a product of two single-particle hopping terms in an ordinary background electromagnetic field $\mathcal{A}$. 
For instance, by resolving (\ref{eqn:dipole_hopping0}) as a product of two terms that hop single electrons along $\hat{x}$ we could write:
\begin{equation}
\begin{split}
    &c^\dagger_{\textbf{r}+\hat{x}} c_{\textbf{r}} c^\dagger_{\textbf{r}+\hat{y}} c_{\textbf{r}+\hat{x}+\hat{y}}\\
    &\to\left(\text{e}^{-i\frac{e}{\hbar}\int_{\textbf{r}}^{\textbf{r}+\hat{x}}\mathcal{A}}c^\dagger_{\textbf{r}+\hat{x}}c_{\textbf{r}}\right)\left(\text{e}^{i\frac{e}{\hbar}\int_{\textbf{r}+\hat{y}}^{\textbf{r}+\hat{x}+\hat{y}}\mathcal{A}}c^\dagger_{\textbf{r}+\hat{y}}c_{\textbf{r}+\hat{x}+\hat{y}}\right)\\
    &=\text{e}^{i\left(A_x(\textbf{r}+\hat{y})-A_x(\textbf{r})\right)}c^\dagger_{\textbf{r}+\hat{x}} c_{\textbf{r}} c^\dagger_{\textbf{r}+\hat{y}} c_{\textbf{r}+\hat{x}+\hat{y}}.
    \label{eq:ringsplit}
    \end{split}
\end{equation} We could have alternatively resolved the ring exchange term as electrons hopping along $y$ which would have resulted in a phase factor of $\text{e}^{i\left(A_y(\textbf{r}+\hat{x})-A_y(\textbf{r})\right)}.$ Hence, if $\mathcal{A}$ is uniform the phase factors vanish and there is no coupling (and thus no polarization response to an applied electric field which is consistent with a dipole conserving system).

Instead of using the conventional electromagnetic vector potential we can take the approach of Ref. \onlinecite{ybh2019}, and introduce a new type of Peierls factor (c.f. (\ref{eqn:Peierls_tb})) that couples to the ring-exchange terms as:
\begin{equation}
\label{eq:quad_phase_factor}
    c^\dagger_{\textbf{r}+\hat{x}} c_{\textbf{r}} c^\dagger_{\textbf{r}+\hat{y}} c_{\textbf{r}+\hat{x}+\hat{y}}\to\text{e}^{iA_{xy}}c^\dagger_{\textbf{r}+\hat{x}} c_{\textbf{r}} c^\dagger_{\textbf{r}+\hat{y}} c_{\textbf{r}+\hat{x}+\hat{y}}.
\end{equation} The Peierls factor depends on a field $A_{xy}$ that we will call a rank-2 gauge field. A single ring-exchange term can be viewed as an operator that either hops an $x$-dipole in the $\hat{y}$-direction or as a $y$-dipole along $\hat{x}.$ That ambivalence is captured by the symmetry of the rank-2 gauge field: $A_{xy}=A_{yx}$.\footnote{We note that $U(1)$ subsystem symmetry, which imposes charge conservation along rows and columns, also prohibits hopping terms that move dipoles longitudinally (parallel to their dipole moment), meaning that our systems of interest couple only to off-diagonal entries of the $A_{ij}$ field.
} 

Now that we have a new space of gauge fields we can imagine making a uniform phase shift of this gauge field across the whole lattice by the  constant value $\mathfrak{q}$:
\begin{equation}
    A_{xy}\to A_{xy}+\mathfrak{q},
\end{equation}
and then study the adiabatic evolution of the Hamiltonian as we slowly drive parameter $\mathfrak{q}$ from $0$ to $2\pi/N_xN_y$ over the time period $T$. In Ref. \onlinecite{dmmh2019dipole} it was shown that such an adiabatic evolution for a charge and dipole insulating system modifies the ground state wave function by a phase that is proportional to the overall quadrupole moment $Q_{xy}$ of the ground state. The Berry phase computed in the parameter space spanned by $\mathfrak{q}$ is:
\begin{equation}
    \gamma_{\mathcal{Q}}=i\int_{0}^{2\pi/N_x N_y}d\mathfrak{q}\langle \Psi_0(\mathfrak{q})\vert\partial_{\mathfrak{q}}\vert\Psi_0(\mathfrak{q})\rangle,
\end{equation} and one can argue similar to Eq. \ref{eq:bphasephys} that it is equivalent to\cite{wheeler2018many,kang2018many,dmmh2019dipole}:
\begin{equation}
\begin{split}
    \gamma_{\mathcal{Q}}=\frac{1}{e}\int_0^T dt\ Q_{xy}\partial_t A_{xy}(t)=\frac{2\pi Q_{xy}}{e L_x L_y},
    \end{split}
    \label{eqn:qp_berry}
\end{equation}\noindent where $Q_{xy}$ is the quadrupole moment and $\partial_t A_{xy}$ is the relevant piece of the rank-2 electric field $E_{xy}.$
This approach was shown to give the correct value of $Q_{xy}$, for the ground state of the ring-exchange model (\ref{eqn:ring-exchange}) where we have:
\begin{equation}
    \gamma_{\mathcal{Q}}=\begin{cases} \pi& \text{when}\ \lambda=1,\ t=0\\ 0& \text{when}\ \lambda=0,\ t=1 \end{cases},
\end{equation} as anticipated from our earlier arguments.

These results provide a powerful framework in which one can calculate many-body quadrupole moments, but the question whether the quantity $\gamma_{\mathcal{Q}}$ is in some sense topologically quantized generally, say in the presence of some protective symmetries, remains nebulous. Some arguments for quantization were given in Refs. \onlinecite{ybh2019,dmmh2019dipole}, but there are still some open questions. To motivate some of these questions let us look closer at the gauge field $A_{xy}.$

To provide a heuristic interpretation of the phase $A_{xy}$ 
 we can find a relationship between the rank-2 and rank-1 gauge fields motivated by Eq. \ref{eq:ringsplit}: $A_{xy}=\partial_y A_x$. Equivalently, by resolving a ring exchange term (\ref{eqn:dipole_hopping0}) as a product of two electron hopping operators along $\hat{y}$, we obtain a different relationship: $A_{xy}=\partial_x A_y$. However, in the absence of the $B_z$ component of the usual rank-1 magnetic field we find that the two equations are equivalent, as $\partial_x A_{y}=\partial_y A_x$; in principle, all of the normalized linear combinations of the two gradients of rank-1 fields are equivalent. For instance, one can pick a symmetric combination\cite{ybh2019,dmmh2019dipole}:
\begin{equation}
    A_{xy}=\frac12\left(\partial_x A_y + \partial_y A_x\right).
    \label{eqn:rank2-rank1}
\end{equation}
Under this identification  $A_{xy}$ inherits its gauge structure directly from the electromagnetic vector potential. The phases $A_i$ transform under the $U(1)$ gauge group as $A_i\to A_i+\partial_i f$ leading to the following rank-2 gauge transformations:
\begin{equation}
    A_{xy}\to A_{xy}+\partial_x\partial_y f.
    \label{eqn:rank2_gauge_transform}
\end{equation}
Rank-2 field theories with these gauge transformations, dubbed ``scalar-charge"\cite{rasmussen16} fracton theories, have attracted a considerable amount of attention in recent years.\cite{pretko2017,pretko2018,pretko2018gauge} 
Particularly, one of the features of scalar-charge fracton field theories is that they exhibit conservation of the dipole moment, which makes them a natural candidate for a field to which dipoles can couple, as we have essentially exploited.

We can make some additional important observations. We have noted that the ring-exchange model is not sensitive to uniform rank-1 fields. Thus, the model is not affected by some configurations of the electromagnetic field $A_{\mu}$ that are \emph{not} pure gauge configurations, and should normally be treated as non-vanishing physical electromagnetic fields. For instance, consider a constant electric field $\textbf{E}=(E_x,0)^T$ generated by $\mathcal{A}_{x}=-E_x t$. Clearly, this rank-1 field configuration leaves $A_{xy}$ given by (\ref{eqn:rank2-rank1}) vanishing, but generates a physical electric field. 
This simply mirrors the fact, that the matter in our theory, instead of carrying electric charges, carries dipole moments which are not affected by constant electric fields.  These observations allow for a wider class of effective ``gauge" transformations of the rank-1 field $A_\mu.$  Picking a particular map from rank-1 to rank-2 fields as in Eq. \ref{eqn:rank2-rank1}, we can see that the spatial components $A_i$ admit the following transformations that leave the rank-2 field invariant up to a rank-2 gauge transformation (\ref{eqn:rank2_gauge_transform}):
\begin{equation}
    A_i\to A_i+\lambda_i,\ \text{where}\  \partial_i\lambda_j+\partial_j\lambda_i=2\partial_i\partial_j f,
    \label{eqn:rank-2_gauge}
\end{equation}
where $i\neq j,$ and $f$ can be an arbitrary function of the spacetime coordinates.
Effectively, the subsystem symmetry, and hence dipole conservation, allows one to enlarge the gauge group for the spatial components of the underlying electromagnetic field. What is missing here is a more unified way to treat these subtle issues.

Let us briefly turn to a related problem. In the two insulating phases of the quadrupolar ring-exchange model, the quadrupole moment is quantized.\cite{dmmh2019dipole,ybh2019} Thus, it is natural to search for a topological invariant that captures this quantized observable. To provide a formulation of the quadrupole Berry phase in terms of a quantized topological invariant it is tempting to rewrite the $A_{xy}$ field in terms of the gradients of rank-1 field, for example, as in Eq. \ref{eqn:rank2-rank1}.
However, this interpretation immediately encounters a problem: it is impossible to find a configuration of the electromagnetic field $A_{\mu}$ that will result in a \emph{uniform} rank-2 field in a periodic system without simultaneously threading some amount of spurious electromagnetic flux through the plane of the periodic lattice. Hence, connecting the rank-2 field to a quantized topological response raises some questions.

In the following sections we present a framework that, instead of connecting the rank-2 field to a non-uniform electromagnetic field,  allows one to alternatively relate the rank-2 field $A_{xy}$ to a Peierls' phase factor derived from a background continuum 2-form field $\mathcal{B}$ that is introduced as a result of gauging global 1-form symmetries. This will allow us to understand $\gamma_{\mathcal{Q}}$ as a phase the ground state wave-function obtains as we interpolate between two configurations of the background \emph{2-form} field that differ by a large gauge transformation. This will give us a precise sense in which we can interpret the properties of $\gamma_{\mathcal{Q}},$ and its symmetry-enforced quantization.

Our construction replaces the subsystem $U(1)$ symmetries with 1-form electric symmetries, and introduces a background field $\mathcal{B},$ that, instead of coupling to a conventional charge current, minimally couples to the electromagnetic field strength $\mathcal{F}.$  As we will see, from this point of view, the Peierls' phase factor attached to the lattice ring-exchange term  (\ref{eq:quad_phase_factor}) does not arise from the pair of charges composing an elementary dipole (which couple to $\mathcal{A}$), but instead from the Wilson line that stretches between these two charges (which couples to $\mathcal{B}$). Hence, we introduce a generalized Peierls' substitution where $A_{xy}$ is interpreted as the lattice field that arises from the background 2-form field $\mathcal{B}$ integrated over a plaquette $p_{xy},$ instead of arising from differences in the rank-1 field $\mathcal{A}.$ That is, the rank-2 field variable can be associated with:
\begin{equation}
    A_{xy}=\frac{e}{\hbar}\iint_{p_{xy}} \mathcal{B},
    \label{eqn:gen_p1}
\end{equation}\noindent which is one of the main conceptual developments of this article. Before we arrive at this result, and the subsequent applications, we first need to introduce and review the idea of higher form symmetries.

\section{Higher-Form Symmetries and Multipole Conservation}
\label{sec:review_gauging}
In this section we offer a brief review of higher-form symmetries where the goal is to appropriately generalize the usual notion of symmetry to extended objects, which for our case are electric flux lines.  We will closely follow Ref. \onlinecite{Gaiotto2015}, but we will specialize to (2+1)d.
Hereafter, we will make a common abuse of language and use ``gauging a symmetry'' to mean ``promoting a global symmetry to a local symmetry.'' 

In the overwhelming majority of theories, global symmetries are unitary operators that act on point-like objects, e.g., charged point-particles.
Nevertheless, this can be generalized to include symmetries acting on extended objects, such as lines or membranes\cite{Gaiotto2015}, however, doing so requires a careful treatment of how, and importantly \textit{where}, a unitary operator acts. 
To treat such symmetries several definitions are in order. First, a symmetry that acts on an $n$-dimensional object is termed an ``$n$-form symmetry." 
Second, objects are ``charged under a symmetry'' if they transform non-trivially under a given symmetry. 
And third, each $n$-form symmetry leads to a conserved $(n+1)$-form Noether current $j,$ which we can use to define the charge operator acting on a \emph{spatial} (sub-)manifold $\mathcal{M}^{d-n}$ in a $d$-dimensional space:
\begin{equation}
    Q(\mathcal{M}^{d-n})=\int_{\mathcal{M}^{d-n}}\star j.
\end{equation}
The key point here is that the $n$-form symmetry action is associated with a manifold $\mathcal{M}^{d-n}$.  Symmetry operators take this manifold as an input to describe where to act. We will consider global symmetry operators that act on a state only at a fixed time, and so the manifold $\mathcal{M}^{d-n}$ in our discussion will always be entirely spatial.

\subsection{0-form Symmetries and Charge Conservation}
The most familiar example of a global symmetry is the global $U(1)$ charge symmetry. This  is a 0-form symmetry that acts by rotating the phase of every point-particle by the same amount at a particular time. In this case  the $(d-0)$-dimensional manifold $\mathcal{M}^d$ is just the entire space. More specifically, let us examine the case of the 0-form $U(1)$ symmetry in QED$_3$.  
The charged objects (electrons) are point-like in space, and the Noether current is given by the electric current 1-form $j=-\rho dt+\vec{j}\cdot d\vec{r}.$
The corresponding charge operator is then associated with the entire 2-dimensional space:
\begin{equation}
    Q(\mathcal{M}^2)=\int_{\mathcal{M}^2} \star j=\int_{\mathcal{M}^3}\rho \;dx dy .
    \label{eqn:noether_charge_0form}
\end{equation}
A representative symmetry operator with parameter $g$ is thus:
\begin{equation}
    U_g(\mathcal{M}^2) = \exp{\left(i g \int_{\mathcal{M}^2}{\star j}\right)}.
    \label{eqn:global_charge_u1}
\end{equation}
This operator acts to rotate the phase of all charged operators (electrons) that intersect (i.e., are within) the spatial 2-volume $\mathcal{M}^2$.  The (3d) Hodge star operation takes the ordinary current 1-form $j$ to a 2-form, and we see $(\star j)_{xy} = j_0$ is the charge density.  Thus, under the action of $U_g(\mathcal{M}^2)$, all electrons inside $\mathcal{M}^2$ will obtain a phase $g$ times their charge.

This operator action is a global symmetry when $g$ is constant in space and $\mathcal{M}^2$ is space itself. Gauging this symmetry means that $g(x)$ can depend on the spatial coordinates, and parameterizes a local phase rotation via
\begin{equation}
    U_g(\mathcal{M}^2) = \exp{\left(i \int_{\mathcal{M}^2}{\star j}\,g(x)\right)}.
    \label{eqn:0_form_symmetry}
\end{equation}
It also comes with the introduction of the usual electric field operator $E_i,$ and the Gauss' law constraint that physical states satisfy:
\begin{equation}
    \left(\sum_{i=1}^d\partial_i E_i - j_0\right)\left|\text{Phys}\right\rangle = 0.
\end{equation}\noindent This relationship enforces the local conservation of charge, and, along with the canonical commutation relations $[A_j(\textbf{r}),E_k(\textbf{r}')] = i\delta_{jk}\delta(\textbf{r}-\textbf{r}')$, defines a gauge transformation on $A_j$:
\begin{equation}
    A_j \rightarrow A_j + \partial_j g(x).
\end{equation}
 As a simple example we see that for $N$ charged point-particles, each located at a point $\textbf{r}_i,$ and carrying a charge $q_i,$ we have $j_0 = \sum_{i=1}^{N}q_i\delta(\textbf{r}-\textbf{r}_i)$. 
The gauged symmetry operator (\ref{eqn:0_form_symmetry}) generates the local phase rotation for a wave-function of $N$ particles located in $\mathcal{M}^2$:
\begin{equation}
    U_g\Psi(\textbf{r}_1,\textbf{r}_2,...,\textbf{r}_N) = \text{e}^{i\sum_{j=1}^N q_j g(\textbf{r}_j)}\Psi(\textbf{r}_1,\textbf{r}_2,...,\textbf{r}_N).
\end{equation}

\subsection{1-form Symmetries and Dipole Conservation}
\label{ssec:1form-polarization}
\begin{figure}
    \includegraphics[width=0.5\textwidth]{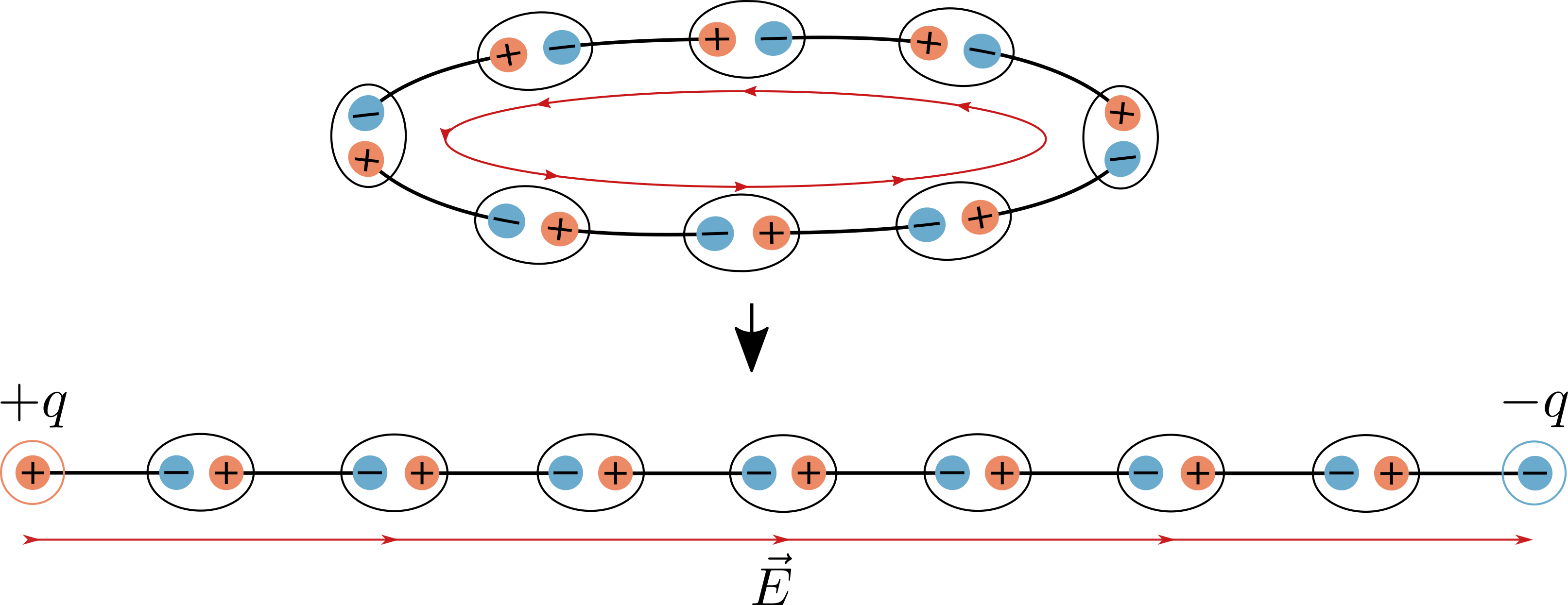}
    \caption{A uniformly electrically-neutral chain carrying a polarization and an associated electric line which can be revealed by breaking the periodic boundary conditions.}
\label{fig:pol_chain1d}
\end{figure}

Physically, a natural starting point for dipole conservation is to require that the system does not conduct electric charge:
\begin{equation}
    \frac{dP_i}{dt}=j_i\equiv 0,
   \end{equation} where $P_i$ is the electric polarization vector. However, we will now motivate that a natural formal setting for systems with conserved dipole moments are models manifesting 1-form symmetries. For the total dipole moment to be a well-defined physical quantity, the total electric charge of the system must vanish. 
   However, a neutral background can still support local dipole moments in the form of closed dipole lines, or open dipole lines with end points on the boundaries (see Fig. \ref{fig:pol_chain1d}).

A key characteristic of charged particles coupled to an electromagnetic gauge field is that electric field lines naturally arise along with the introduction of a dipole moment. To see this, consider a simple electron hopping operator $c^\dagger_{\textbf{r}+a\hat{x}}c_{\textbf{r}}$. When acting on a neutral background this operator annihilates one electron at a site $\textbf{r}$ leaving a positively charged hole, and creates an electron with negative charge  at a site with coordinates $\textbf{r}+a\hat{x}$, thus effectively creating an elementary dipole moment $\textbf{d}=-ea\hat{x}$. Importantly, in a background electromagnetic field, this simple hopping operator needs to be modified by a Peierls' phase\cite{Peierls1933} $A_x(\textbf{r})$ defined on the link that stretches between the two aforementioned sites:
\begin{equation}
    c^\dagger_{\textbf{r}+a\hat{x}}c_{\textbf{r}}\to \text{e}^{iA_x(\textbf{r})}c^\dagger_{\textbf{r}+a\hat{x}}c_{\textbf{r}}.
    \label{eqn:Peierls_sub}
\end{equation}
Upon quantization of electromagnetic field on the lattice, we impose the following commutation relations on a lattice vector potential $\hat{A}_{l}$ (which includes in its definition the factor of $e/\hbar$) and an electric field $\hat{E}_{l'}$ (which is defined in the units of $-e/\varepsilon_0$)  that live on the links $l$ and $l'$ respectively:
\begin{equation}
    [\hat{A}_{l},\hat{E}_{l'}]=i\delta_{l,l'}.
    \label{eqn:bos_ladder}
\end{equation}
Thus, we see that the Peierls phase introduced in Eq. (\ref{eqn:Peierls_sub}) acts as a ladder operator for the electric field since we have:
\begin{equation}
    \left[\hat{E}_l,\text{e}^{\pm i\hat{A}_l}\right]=\pm\text{e}^{\pm i\hat{A}_l}.
    \label{eqn:electric_commutator}
\end{equation}
Hence, we can regard $\text{e}^{\pm i\hat{A}_l}$ as a creation/annihilation operator, and $\hat{E}_l$ as a number operator, for a bosonic degree of freedom on the link $l$ representing a line of electric flux.
Thus, we see that the hopping operator (\ref{eqn:Peierls_sub}) not only accounts for the changes in local electric charges, but it also naturally takes into account the change in the electric flux that passes between the pair of sites $\textbf{r}$ and $\textbf{r}+a\hat{x}$ (see lower part of Fig. \ref{fig:pol_chain1d}).

From this observation, and the constraints implied by the absence of free charges/charged excitations, one can calculate the bulk polarization of a system by counting the associated oriented electric flux lines as depicted in red in Fig. \ref{fig:pol_chain1d}.  
To provide some physical intuition, let us focus on the single polarized line depicted in Fig. \ref{fig:pol_chain1d}.
The closed polarized line preserves the charge-neutrality at every point of the system, but terminating this line on a boundary creates a pair of opposite charges at the two different boundary points. This accompanies an electric field configuration with the total amount of electric flux equal to $\Phi^x_E= q/\varepsilon_0$, where $\varepsilon_0$ is the vacuum permittivity. 
Furthermore, in addition to introducing the electric flux to the bulk of the system, it reflects a polarization of $p_x=-q$. In other words, we can employ the electric flux going through the bulk of the polarized material to calculate the polarization $p_x=-\varepsilon_0 \Phi^x_{E}$.

Now we will show how one can calculate the polarization using the associated electric flux instead of charge density, and how this is connected to a 1-form symmetry operator. Consider a (2+1)d system, and  impose periodic boundary conditions in both spatial directions. This system has no electric charges but can still support closed polarization lines, i.e., closed electric flux lines.  
To calculate the polarization along, say, the $\hat{x}$-direction in the completely periodic system, we simply need to calculate the net electric flux winding around in that direction. 
Formally, this can be done by integrating the electric field passing through a one-dimensional surface $\mathcal{M}^1_x$ winding along $\hat{y}$. 
For example, we may take $\mathcal{M}^1_x$ to be defined by the equation $x=x_0,$ and then calculate:
\begin{equation}
    \Phi_{E}^x=\int_{\mathcal{M}_x^1}\vec{E}\cdot \vec{n} dy=\int_{\mathcal{M}_x^1}E_x dy = c\int_{\mathcal{M}_x^1}\star\mathcal{F},
    \label{eqn:electric_integral}
\end{equation}
where $\vec{E}$ is the electric field, $\vec{n}$ is the unit normal vector to $\mathcal{M}_x^{1},$ and $\mathcal{F}$ is the electromagnetic field strength. 
We can then identify the polarization $p_x=-\varepsilon_0 \Phi_{E}^x.$

Importantly, there is an alternative interpretation of the last integral in Eq. (\ref{eqn:electric_integral}).
Just as the integral of $\star j$ is the generator of the global $U(1)$ charge conservation symmetry, the integral of $\star\mathcal{F}$ can be understood as a generator of an \emph{electric 1-form $U(1)$ symmetry} defined over $\mathcal{M}_x^1$:
\begin{equation}
    U_g(\mathcal{M}_x^1)=\exp\left(igc\int_{\mathcal{M}_x^1}\star\mathcal{F}\right).
    \label{eqn:1form_Ug}
\end{equation}
Conservation of the total electric charge is equivalent to demanding the invariance with respect to global $U(1)$ phase rotations (\ref{eqn:global_charge_u1}), and analogously, demanding that the system is invariant with respect to these 1-form transformations defined over various closed surfaces $\mathcal{M}^1$ leads to the conservation of the electric flux in the system, which is, in our case, equivalent to the polarization. This key observation allows us to reformulate the notion of a dipole-conserving system as a system that is invariant with respect to 1-form electric $U(1)$ transformations, and is a key concept in this article. 

In this formalism, different components of the polarization $p_i$ are associated to conserved charges of electric 1-form symmetries defined over closed $(d-1)$-dimensional manifolds $\mathcal{M}^{d-1}_{i}$ that are periodic along the $(d-1)$ transverse directions that are orthogonal to $\hat{r}_i$. 
Explicitly, we can calculate the electric polarization $p_i$ along the $\hat{x}_i$ direction with the help of the electric 1-form symmetry operator (\ref{eqn:1form_Ug}): 
\begin{equation}
    p_i=-\frac{eL_i}{2\pi V}\text{Im} \log\left\langle U_{2\pi \varepsilon_0/e}(\mathcal{M}_i^{d-1})\right\rangle,
    \label{eqn:Pol-1form}
\end{equation} where $L_i$ is the system size along $\hat{x}_i,$ and $V$ is the volume of the system.

This formula for the polarization is similar to they many-body operator formalism introduced by Resta to calculate the polarization\cite{resta1998}. To make the connection more clear it is instructive to witness how this formula applies in the context of $(1+1)d$ electrodynamics with a $\theta$-term. In the presence of inversion symmetry the value of $\theta$ is quantized to $0,\pi,$ and it captures the response properties of the $(1+1)d$ topological polarized insulator discussed earlier. Consider the following Lagrangian density for electrodynamics on a circle:
\begin{equation}
    \mathcal{L}=-\frac{1}{2\mu_0 c}\mathcal{F}\wedge\star \mathcal{F}+\frac{e\theta}{2\pi}d\mathcal{A}=\left(-\frac{\varepsilon_0}{2}E_x^2+\frac{e\theta}{2\pi}E_x\right)dtdx,
\end{equation}
where we substituted $E_x=c\left(\partial_t\mathcal{A}_x/c-\partial_x\mathcal{A}_t\right)$. Now we would like to quantize this theory on a 1-dimensional chain with $N$ sites and the same number of links.
The canonical momentum to $\mathcal{A}_x$ is:
\begin{equation}
    \pi_x=\frac{\partial\mathcal{L}}{\partial\dot{\mathcal{A}}_x}=-\varepsilon_0 E_x+\frac{e\theta}{2\pi}.
\end{equation}
 We need to discretize this theory by introducing lattice variables on the links $l$: $A_l\equiv\frac{e}{\hbar}\int_l \hat{\mathcal{A}}_x(x)dx$ (which is $2\pi$ periodic) and $\hat{\pi}_l\equiv\int_l \hat{\pi}_x(x)dx$. Now we can obtain the following lattice Hamiltonian:
\begin{equation}
    \hat{H}=\frac{e^2a}{2\varepsilon_0}\sum_{l=1}^N\left(\frac{\hat{\pi}_l}{ea}-\frac{\theta}{2\pi}\right)^2,
    \label{eqn:1+1d_Ham}
\end{equation} where $a$ is the lattice constant.
Imposing canonical commutation relations produces
\begin{equation}
[\hat{A}_l,\hat{\pi}_{l'}]=iea\delta_{l,l'}.
\label{eqn:comm_rel1d}
\end{equation}
We should also restrict the Hilbert space to the space of physical gauge-invariant states by imposing the Gauss' constraint:
\begin{equation}
    \partial_x E_x(x)\ket{\text{Phys}}=0 \leftrightarrow (\hat{E}_{l+1}-\hat{E}_l)\ket{\text{Phys}}=0,\ \forall\ l.
    \label{eqn:Gauss_constr}
\end{equation}
For a one-dimensional chain this constraint requires that for any physical state, the amount of electric flux (quantified by the integer $n$) is the same for every link of the lattice. Such states of this Hamiltonian are given by:
\begin{equation}
    \ket{\psi_n}=\text{e}^{in\sum_l \hat{A}_l}\ket{\text{vac}},\ n\in\mathbb{Z}.
\end{equation}
Taking into account the commutation relationship (\ref{eqn:comm_rel1d}) we can now write down the spectrum of the Hamiltonian (\ref{eqn:1+1d_Ham}):
\begin{equation}
    \mathcal{E}_n=\frac{e^2 L}{2\varepsilon_0 c^2}\left(n-\frac{\theta}{2\pi}\right)^2,
\end{equation}
where $L=Na$. 

Consider now the special case when $\theta=\pi$ which we expect to correspond to a polarization $p_x=\pm e/2$. For this value of $\theta$, the energy spectrum becomes doubly-degenerate. Let us consider the two lowest energy states $\ket{\psi_0}$ and $\ket{\psi_1}$ that form the ground state subspace, and then calculate their polarization via Eq. \ref{eqn:Pol-1form}. Since these states are eigenstates of the canonical momentum we can easily calculate the expectation value of the electric field. For $\hat{\pi}_l$ on a link $l$ we find:
\begin{equation}
\begin{split}
    &\bra{\psi_0}\hat{\pi}_l\ket{\psi_0}=-a\varepsilon_0\langle \hat{E}_l\rangle_0+\frac{ea}{2}=0\rightarrow \langle \hat{E}_l\rangle_0 =\frac{e}{2\varepsilon_0},\\
    &\bra{\psi_1}\hat{\pi}_l\ket{\psi_1}=-a\varepsilon_0\langle \hat{E}_l\rangle_1+\frac{ea}{2}=ea\rightarrow \langle \hat{E}_l\rangle_1 =-\frac{e}{2\varepsilon_0}.
\end{split}
\label{eqn:E_expect}
\end{equation} 
Due to the Gauss' constraint (\ref{eqn:Gauss_constr}), the expectation value of the electric field operators are the same on all links of the chain for any given state.
We can now apply Eq. \ref{eqn:Pol-1form} to determine the electric polarization for these two states. To do so, we need to calculate the expectation value of the operator:
\begin{equation}
    U_{2\pi\varepsilon_0/e}(\mathcal{M}^0_l)=\exp\left(2\pi i\frac{\varepsilon_0}{e}\hat{E}_l\right),
    \label{eqn:quantized_1d_lattice_op}
\end{equation}
where $\mathcal{M}^0_l$ in this case denotes a particular link of the chain.
Using the expressions in Eq. (\ref{eqn:E_expect}), we obtain the following values of polarization for states $\ket{\psi_0}$ and $\ket{\psi_1}$:
\begin{equation}
    p^0_x=\frac{e}{2}\mod e,\ p^1_x=-\frac{e}{2}\mod e.
\end{equation}
In other words, we find that $p_x^0=p_x^1=e/2\mod e$.
This result could have also been calculated from the  Goldstone-Wilczek response term (\ref{eq:polar_resp0})\cite{Goldstone81}.

Now, given the simple context of the one-dimensional lattice, let us examine more closely the 1-form symmetry operator. 
In a quantized electromagnetic theory, the operator (\ref{eqn:quantized_1d_lattice_op}) acts by shifting one of the link variables $\hat{A}_l$ by $2\pi$:
\begin{equation}
    \text{e}^{2\pi i\frac{\varepsilon_0}{e}\hat{E}_l}\hat{A}_{l'}\text{e}^{-2\pi i\frac{\varepsilon_0}{e}\hat{E}_l}=\hat{A}_{l'}-2\pi\delta_{l,l'}.
\end{equation}
In a lattice gauge theory, such a shift of the lattice variable performs a large gauge transformation, i.e., $\sum_{l=1}^{N}\delta A_l=0\mod 2\pi$. 
This shift is also gauge-equivalent to the following uniform shift across all lattice variables:
\begin{equation}
    A_l\to A_l-\frac{2\pi}{N}.
\end{equation}
This uniform shift can be implemented by the operator:
\begin{equation}
    \tilde{U}_{2\pi\varepsilon_0/e}=\exp\left(\frac{2\pi i\varepsilon}{Ne}\sum_{l=1}^N\hat{E}_l\right),
\end{equation}
which is a gauge-equivalent to Eq. \ref{eqn:quantized_1d_lattice_op}.\footnote{Note that due to the Gauss' law constraint, these operators evaluate to the same quantity on all physical states. Since both of these operators implement the same large gauge transformation, we can think of them as being gauge-equivalent to each other differing by the (small) gauge transformation $A_x\to A_x+\partial_x f(x)$ implemented by $f(x)=-2\pi x/L+2\pi\Theta(x-x_0)$, where $\Theta(x)$ is a Heaviside step function and the coordinate $x_0$ is located on the link $l$. In terms of the quantized electromagnetic field operators, this small gauge transformation is implemented by the operator $(U_{2\pi\varepsilon_0/e})^{-1}\tilde{U}_{2\pi\varepsilon_0/e}$.} It is well-known\cite{resta1998,oshikawa2000}, that this particular large gauge transformation can be implemented by the following twist operator acting on the fermionic degrees of freedom:
\begin{equation}
    U_X=\exp\left(-\frac{2\pi i}{L}\sum_{i=1}^{N}x_i\hat{n}_i\right),
\end{equation}
where $x_i$ is the coordinate of the lattice site $i,$ and $\hat{n}_i$ is the particle number operator on that site. 
It is no surprise then that we can substitute $U_X$ for $U_{2\pi\varepsilon_0/e}(\mathcal{M}^0_l)$ in Eq. \ref{eqn:Pol-1form} to obtain the well-known result:
\begin{equation}
    p_x=-\frac{e}{2\pi}\text{Im} \log\left\langle U_X\right\rangle.
    \label{eqn:Pol-1form_UX}
\end{equation}
Intuitively, this equivalence can also be explained by noticing that the quantum-mechanical electric field operator in a time-independent system is just the position operator $\hat{E}=-\frac{e}{\varepsilon_0}\hat{X}$, where $\hat{X}=\frac{1}{L}\sum_i x_i \hat{n}_i$. 

In summary, we have presented an alternative way to calculate the charge polarization by counting the electric flux in a system. While this may seem like a formal development, our perspective will enable us to make advances that were not manifest without it.

\section{Peierls Substitution for Dipoles}\label{sec:peierls}
Having an interpretation for dipole moments as electric flux lines allows us to consider an alternative way to couple dipoles to a gauge field. The global $U(1)$ 1-form symmetries,
\begin{equation}
    U_g(\mathcal{M}^1)=\exp\left(igc\int_{\mathcal{M}^1}\star\mathcal{F}\right)
  \end{equation}\noindent can be gauged  by letting $g(\textbf{r})$ be spatially dependent.  For an ordinary 0-form (gauge) symmetry in $d+1$ spacetime, an operator that acts at a specific time can have $g(\textbf{r})$ vary in $d$-dimensional space.
Let us now illustrate how this generalizes ``local" 1-form symmetries. 

\begin{figure}
    \includegraphics[width=0.24\textwidth]{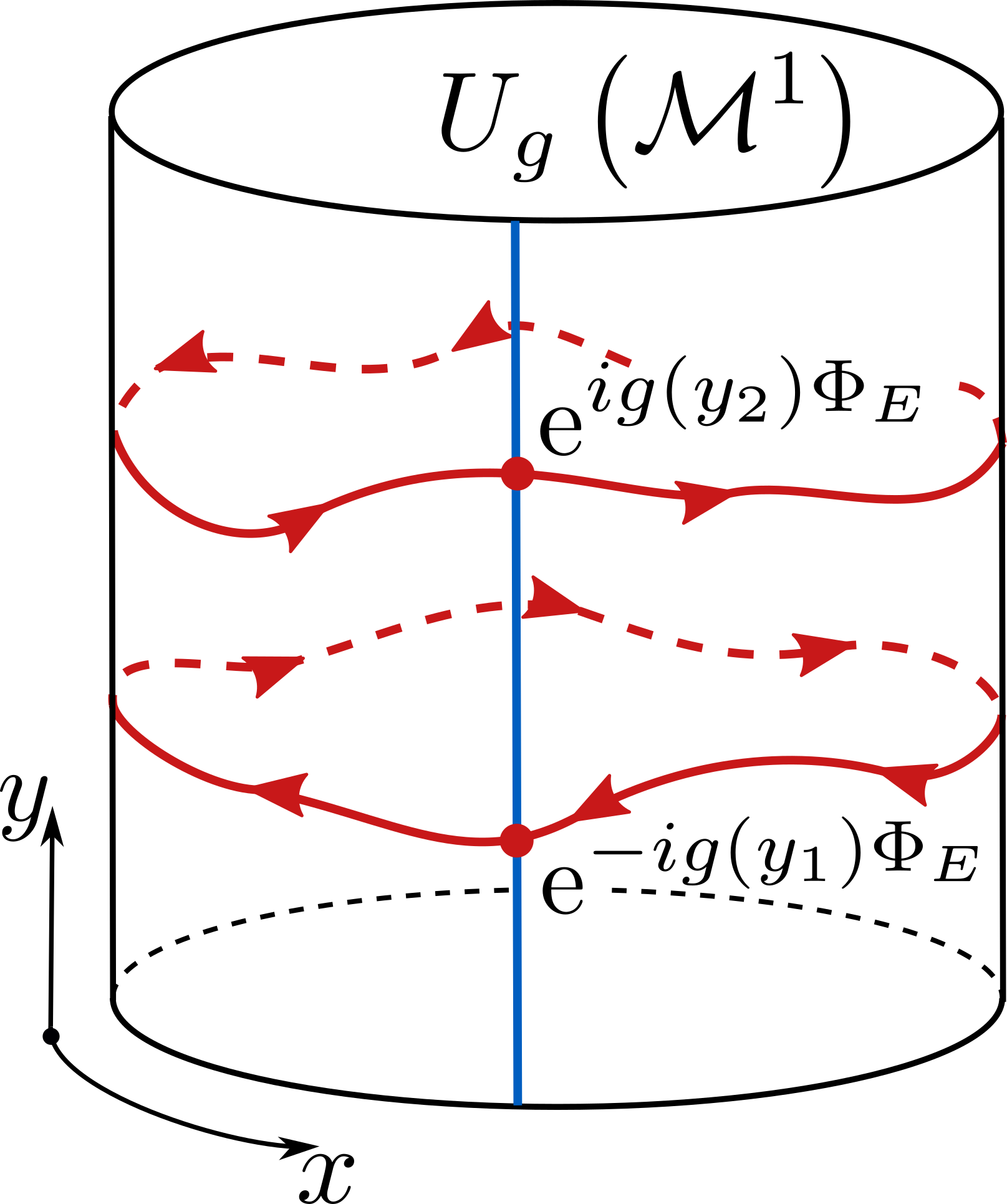}
    \caption{Two opposite Wilson lines carrying electric flux $\Phi_E$ wrapping around $x$-direction may be modified by arbitrary phases under the action of $U_g(\mathcal{M}^1)$ if the parameter $g$ depends on $y$. The manifold $\mathcal{M}^1$ is represented by the blue line.}
    \label{fig:gauge_1form}
\end{figure}

We have shown that global 1-form symmetries are defined over co-dimension 2 spatial manifolds, which for our case of (2+1)d are lines ($\mathcal{M}^1$). The symmetry operator acts on one-dimensional objects that intersect $\mathcal{M}^1$ at a collection of points. 
To illustrate, we show a typical configuration in Fig. \ref{fig:gauge_1form}.  The cylinder is periodic in the $x$-direction, and open in the $y$-direction.  We can choose a symmetry corresponding to the manifold $\mathcal{M}^1$ which is represented by the blue vertical line. The global 1-form symmetry operator $U_g(\mathcal{M}^1)$ will rotate the global phase of a quantum state according to the the total electric flux  $\int_{\mathcal{M}^1}\star\mathcal{F}$. In Fig. \ref{fig:gauge_1form} we depicted a pair of electric flux lines with opposite orientation and the same charge. Hence, the total phase generated by $U_g(\mathcal{M}^1)$ on this state is trivial $\text{e}^{i(g-g)\Phi_E}=1$. 

In general, as we describe in Appendix \ref{app:1formstuff}, because of the commutation relations between the electric field and the gauge field $\mathcal{A},$ a global 1-form symmetry action effectively shifts $\exp\left(i\frac{e}{\hbar}\int_{\Gamma}\mathcal{A}\right)\to \exp\left(i\frac{e}{\hbar}\int_{\Gamma}\mathcal{A}\right)\exp(ig\Phi_{E,\Gamma})$ where $\Gamma$ is any closed Wilson loop that intersects the symmetry manifold $\mathcal{M}^1,$ and $\Phi_{E,\Gamma}$ is the electric flux carried by the Wilson line $\Gamma.$ Hence, an alternative interpretation of $U_g(\mathcal{M}^1)$ is that it effectively threads magnetic flux $\Phi_B=\frac{g h}{e}\Phi_E$ through the center of the cylinder that will affect every Wilson line that encircles the periodic $x$-direction. 

To gauge this symmetry so that it becomes local, we will allow $g$ to vary as a function of the coordinate on $\mathcal{M}^1$, i.e., we take $g=g(y)$ for the configuration in Fig. \ref{fig:gauge_1form}. 
The gauged version of the 1-form symmetry operator $U_g$ is:
\begin{equation}
    U_g(\mathcal{M}^1) = \exp{\left(ic\int_{\mathcal{M}^1}{\star \mathcal{F}\ g(y)}\right)}.
\end{equation} However, in the local case, by letting $g$ be a function of $y$, we are effectively inserting a different amount of magnetic flux through different slices of the cylinder.  If the magnetic field is not allowed to escape through the surface of the cylinder, this situation is physically impossible in the absence of magnetic monopoles.  But as we will now see, the proper way to treat the local symmetry requires the introduction of a new gauge field for the 1-form symmetry that we have so far neglected, and which will compensate for this issue.

\subsection{Introduction of a 2-form gauge field}
\label{sec:2form_description}

Gauging an $n$-form symmetry requires the introduction of an $(n+1)$-form gauge field. 
Thus, when we gauge the electric 1-form symmetry, we will consider a 2-form field\cite{Komargodski19} $\mathcal{B}$.
In Appendix \ref{app:1formstuff} we show that the action of the 1-form symmetry on a Wilson line operator amounts to shifting $\mathcal{A}$ by a generic 1-form $\lambda$:
\begin{equation}
    \mathcal{A}\to\mathcal{A}+\lambda.
\end{equation}
Under these transformations, $\mathcal{F}=d\mathcal{A}$ is no longer invariant, i.e., this process can seemingly generate non-vanishing physical electromagnetic fields. To compensate we couple the system to a 2-form field $\mathcal{B}$ by extending the derivative $d\mathcal{A}\to d\mathcal{A}-\mathcal{B}.$ This is analogous to the minimal coupling of a compact charged scalar field $\varphi\sim \varphi+2\pi$ when gauging electromagnetism: $d\varphi\to d\varphi - \mathcal{A}$.
 Therefore, when treating dipole conserving systems using 1-form symmetries,  the new gauge invariant quantity we need to consider is $\mathcal{F} - \mathcal{B}$. 
 
Let us make a few comments. Under a gauge transformation parameterized by a 1-form field $\lambda$, our fields transform as
\begin{equation}
    \begin{split}
        \mathcal{A} &\rightarrow \mathcal{A} + \lambda\\
    \mathcal{F} &\rightarrow \mathcal{F} + d\lambda\\
    \mathcal{B} &\rightarrow \mathcal{B} + d\lambda.
    \end{split}
    \label{eqn:gen_gauge}
\end{equation}
We see that this transformation \emph{changes} the physical electromagnetic field $\mathcal{F}$ as we mentioned above. Although this may seem unphysical at a first glance, we have already encountered a similar extension of the gauge freedom while reviewing rank-2 gauge fields in Section \ref{sec:review_Berry}. 
There, the additional gauge freedom (\ref{eqn:rank-2_gauge}) was a consequence of matter consisting of dipole moments that are restricted by the $U(1)$ subsystem symmetry to move transversely on rows and columns of the lattice. 
Here, we instead impose 1-form symmetries that are similar, but unlike subsystem symmetries they are not restricted to act only on a certain subset of co-dimension 1 manifolds and so, gauging these symmetries naturally results in a richer gauge group (\ref{eqn:gen_gauge}). Importantly, this gauge structure allows for both  ``small" and ``large" transformations that are distinguished by computing the Chern class of $\lambda$:
\begin{equation}
\label{eqn:2form_lgt}
    \int_{\mathcal{M}^2} d\lambda = 2\pi n,\ n\in\mathbb{Z}.
\end{equation}
When this is nonzero, the transformation is \textit{large}.  Note that the Chern class of $\lambda$ is insensitive to the 1-dimensional topology of $\lambda$ that might support nonzero winding around the non-contractible loops of the spatial manifold;  indeed, as we have mentioned these are just flux insertions  that are equivalent to the \emph{global} 1-form transformations, and are not any type of large 1-form gauge transformatoin. 
 
 \subsection{Dynamics of Electric Lines Coupled to a 2-form Field}\label{ssec:gen_AB}
 To gain some physical intuition about the 2-form gauge field we can consider a Maxwell action modified to exhibit gauge invariance under local 1-form symmetries:
\begin{equation}
    \begin{split}
        \mathcal{L} &= -\frac{1}{2\mu_0c}(\mathcal{F} - \mathcal{B}) \wedge \star (\mathcal{F} - \mathcal{B})\\ 
    &= -\frac{1}{2\mu_0c}\mathcal{F} \wedge \star \mathcal{F} - \frac{1}{2\mu_0c}\mathcal{B} \wedge \star \mathcal{B} + \frac{1}{\mu_0c}\mathcal{B} \wedge \star \mathcal{F}.
    \end{split}
    \label{eqn:lagrangian}
\end{equation}
The last term is a contact term:
\begin{equation}
    \frac{1}{\mu_0c}\mathcal{B} \wedge \star \mathcal{F} = \frac{1}{2\mu_0c} \mathcal{B}_{\mu\nu}\mathcal{F}^{\mu\nu}d^{d+1}x,
    \label{eqn:FB_contact}
\end{equation}
where $\mathcal{B}=\frac12 \mathcal{B}_{\mu\nu}dx^\mu \wedge dx^\nu$ and $\mathcal{F}=\frac12 \mathcal{F}_{\mu\nu}dx^\mu \wedge dx^\nu$. This term is analogous to the usual coupling between a gauge field and a current $\mathcal{A}\wedge\star J = A_\mu J^\mu$, except in this case the gauge field is the 2-form $\mathcal{B}$ and the matter field is the electromagnetic field $\star\mathcal{F}.$  This is important because it indicates how the field $\mathcal{B}$ couples to the electric lines in our system. For example, this contact term identifies $\mathcal{B}_{tx}$ as a kind of scalar potential for non-contractible $x$-electric lines, which are positively charged under 1-form symmetry. This  is similar to how we identify $\mathcal{A}_t$ with the regular electro-static potential for charged particles.

The key feature of this type of theory for our analysis is that the electric lines (i.e., Wilson lines) are charged under the 1-form symmetry and couple to the 2-form gauge field $\mathcal{B}.$  Indeed, a Wilson line is created through the action of $\star \mathcal{F}$ along a loop, thus creating an electric string.  However, as we can see, this process, and the dynamics of electric lines in general, are no longer gauge invariant, because $\mathcal{F}$ transforms under the 1-form symmetry.  To fix this we follow an analogous procedure to gauging a 0-form symmetry.  In the usual case of QED, the electron operator $\bar{\psi}$ is not gauge invariant on its own since it is charged. To construct something gauge invariant we attach this operator  to an oppositely charged $\psi$ via a Wilson line.  Then the whole combination of particle, Wilson line, and antiparticle is gauge invariant.
In the case of a gauged 1-form symmetry this means that a Wilson-line by itself is not gauge invariant. Hence to form something invariant we need to create a pair of oppositely oriented Wilson lines and attach them via a ``Wilson surface.''  Such a pair of Wilson lines will have opposite charges under the electric 1-form $U(1)$, and are an analog of the particle/antiparticle pair.  To establish gauge invariance, the gauge field $\mathcal{B}$ is integrated on this surface in exactly the same way $\mathcal{A}$ is integrated on a line.  This cancels the gauge variance of the Wilson lines, making the combination of $\bar{\psi}\psi,$ Wilson line, oppositely oriented Wilson line, and Wilson surface, gauge invariant under the 1-form symmetry.

\begin{figure}
    \centering
    \includegraphics[width=0.5\textwidth]{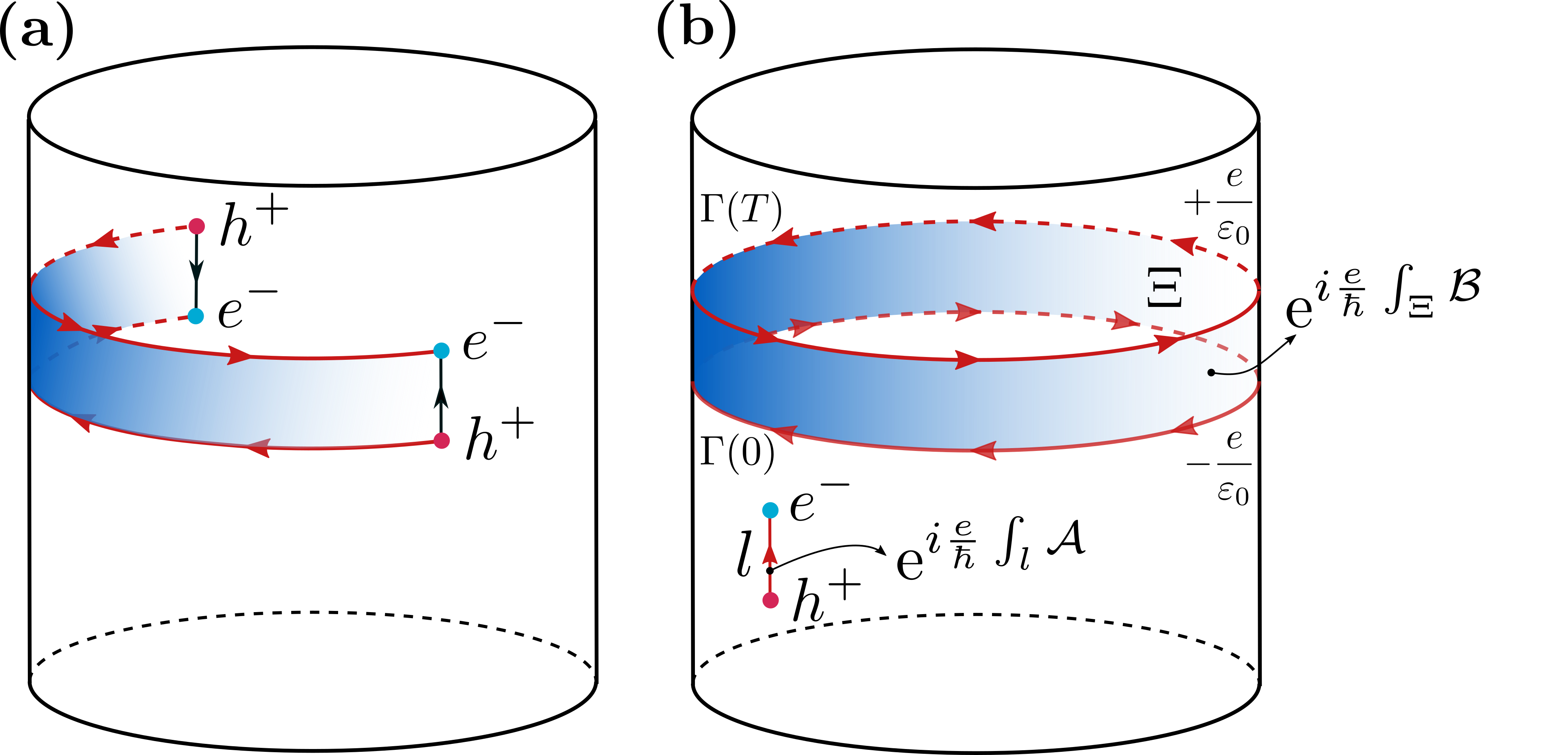}
    \caption{A pair of opposite running electric lines created by a successive application of the dipole hopping operator, that transports $y$-dipole along the $\hat{x}$-direction. In other words, dragging a dipole all around the manifold results in an operator that is a product of two parallel and oppositely running operators.}
    \label{fig:2wilson_lines}
\end{figure}

Let us consider an example to see precisely how this works. The probe 2-form current for a single, dynamical Wilson line defined along a moving path $\Gamma(t)$, parameterized by a variable $s$ along its length, can be written as:
\begin{equation}
\begin{split}
    \mathcal{F}&=\frac{1}{2} \mathcal{F}_{\mu\nu}(\textbf{r})dr^\mu\wedge dr^\nu\\
    &=\frac{q}{2c\varepsilon_0} \delta(\textbf{r} - \textbf{r}')\left(\partial_s r_\mu \partial_{t}r_\nu - \partial_s r_\nu \partial_{t}r_\mu\right)dr^\mu\wedge dr^\nu,
    \end{split}
    \label{eqn:wil_lin_cur}
\end{equation}
where $\textbf{r}'\in\Gamma(t),$ and $\partial_s \textbf{r}$ is a vector that is tangent to $\Gamma(t)$ at a point $\textbf{r}\in\Gamma(t)$. In simpler terms we can consider a charge $q$ particle that one translates around a path $\Gamma(t_0)$ to create the Wilson line configuration at any given time $t_0$. Importantly, this expression only captures the transverse movement of Wilson lines during which a line $\Gamma$ sweeps out a two-dimensional surface.\footnote{Transporting $\Gamma$ along itself without sweeping any surface does not contribute to Eq. (\ref{eqn:wil_lin_cur}), which is similar to how a rotation of a point particle around its center -- a rearrangement of internal degrees of freedom -- is not captured by the expression for a single-particle current.} As a concrete example, let us work in $(2+1)d,$ and let the $\hat{x}$-direction be periodic where $x\sim x+ L_x$. 
Imagine a process in which one creates a closed electric line by winding a charge-$q$ particle in the $\hat{x}$ direction along a path $\Gamma(0)$ as shown in Fig. \ref{fig:2wilson_lines}(b). 
This line is then parallel-transported in the $\hat{y}$-direction over a time period $T$.
The corresponding Wilson line current obtained via Eq. \ref{eqn:wil_lin_cur} is:
\begin{equation}
\label{eq:ws_area}
\begin{split}
    \mathcal{F}=\frac{q}{c\varepsilon_0}\delta(y - y'(t))\left[cdx\wedge dt + \frac{\partial_{t}y}{c}\ dx\wedge dy\right],
    \end{split}
\end{equation}
where the delta function is non-zero only along a path $\Gamma(t)$ which is defined at any given point in time by the equation $y=y'(t)$.

Using this configuration we can now illustrate the physical properties of the coupling between $\mathcal{B}$ and electric flux lines. Let us first verify that this background current indeed carries a fixed amount of electric flux during the whole process.
We do this by counting the electric flux passing through any one-dimensional spatial line $\mathcal{M}^1$ that is parallel to the $y$-axis and oriented in the \emph{positive} $\hat{y}$-direction. To be explicit, consider $\mathcal{M}^1$ defined by the equation $x=x_0$ with the orientation in the positive $\hat{y}$-direction. The electric flux at any point in time $t$ is:
\begin{equation}
    c\int_{\mathcal{M}^1}\star \mathcal{F}=\frac{q}{\varepsilon_0}\int_{x=x_0}dy\ \delta(y-y')=\frac{q}{\varepsilon_0},
\end{equation} where we essentially had to integrate the $\mathcal{F}^{x0}$ component of $\mathcal{F}$ over $\mathcal{M}^1$. 
Now let us compute the phase picked up by our electric line when it sweeps out a tubular surface $\Xi$ and is parallel-transported in a background field $\mathcal{B}$ as shown in Fig. \ref{fig:2wilson_lines}(b). According to the contact term (\ref{eqn:FB_contact}) we have:
\begin{equation}
    \begin{split}
        \Delta\phi = \frac{1}{\hbar\mu_0c}&\int \mathcal{B}\wedge \star \mathcal{F}=\frac{1}{\hbar\mu_0} \int\ dtd\textbf{r}^2 \frac 12 \mathcal{B}^{\mu\nu}\mathcal{F}_{\mu\nu} \\
               &= \frac{q}{\hbar}\int_\Gamma dx \int_0^T dt\ \frac{1}{2}\mathcal{B}^{xy}\partial_t y'= \frac{q}{\hbar}\int_\Xi \mathcal{B},
    \end{split}
    \label{eqn:surf_hol}
\end{equation}
where we used that $\mu_0\varepsilon_0c^2=1$. In the context of differential geometry the phase factor we get from (\ref{eqn:surf_hol}) is known as a surface holonomy\cite{waldorf2006}. 
From this result we immediately see that the background $\mathcal{B}$ field attaches phases to the (transverse) motion of Wilson loops in analogy with how a background $\mathcal{A}$ field attaches phases to moving point-particles. This is a key understanding of how the $\mathcal{B}$ field will couple to matter on the lattice

Before we move on to the implications for lattice models we will briefly mention one more generic result. If we introduce periodic boundary conditions in the $\hat{y}$-direction of the cylinder, then we can transport the Wilson loop around the resulting torus which will generate in a closed Wilson surface. Now, requiring the periodicity of the phase (\ref{eqn:surf_hol}) during this process, we find that the integral of the gauge field  $\mathcal{B}$ around the whole manifold must be quantized. With the periodic boundary conditions in place, the total phase picked up by a Wilson line that gets transported all the way around the torus can take values equal to only $2\pi$ times an integer:
\begin{equation}
    \frac{e}{\hbar}\int_{T^2} \mathcal{B} = 2\pi n,\ n\in\mathbb{Z},
\end{equation}
where $e$ is the electron charge (which also serves serves as an elementary Wilson line charge).
This is exactly analogous to how an integer number of static magnetic flux quanta passing through a one-dimensional ring have no physical consequences for electrons moving around the ring. This integer $n$ cannot be changed by smooth transformations of $\mathcal{B}$ that are connected to the identity.
Performing a gauge transformation of $\mathcal{B}$, we find that the change in the surface holonomy is
\begin{equation}
    \frac{e}{\hbar}\int_{T^2}\mathcal{B} \rightarrow \frac{e}{\hbar}\int_{T^2}\mathcal{B} + \frac{e}{\hbar}\int_{T^2} d\lambda.
\end{equation}
We find that the extra term either vanishes, when $\lambda$ is `small', or, as we saw in Eq. \ref{eqn:2form_lgt}, it is equal to an integer multiple of $2\pi$ when $\lambda$ is `large.' The large gauge transformations thus preserve the boundary conditions for Wilson lines; this is exactly analogous to how the large gauge transformations of $\mathcal{A}$ preserve the boundary conditions for electrons moving around periodic manifold. This quantization condition on the gauge field $\mathcal{B}$ will be relevant for the discussion in Sec. \ref{sec:response}.

\subsection{2-form Peierls Substitution on the Lattice}
\label{sec:2form_Berry_phase_old}
We now return to the question of coupling gauge fields to dipole conserving lattice models, such as the ring-exchange model given in Eq. \ref{eqn:ring-exchange}. Ring-exchange terms naturally conserve dipole moments, and appear in the lattice models of interest in the following form:
\begin{equation}
    c^\dagger_{\textbf{r}} c^{\phantom{\dagger}}_{\textbf{r}+\hat{x}} c^\dagger_{\textbf{r}+\hat{x}+\hat{y}} c^{\phantom{\dagger}}_{\textbf{r}+\hat{y}}.
    \label{eqn:ring-exchange1}
\end{equation}
This operator has a nice, albeit two-fold, physical interpretation: we can think of it either as a term that hops an elementary $x$-dipole one lattice spacing along $\hat{y},$ or as a term that hops and elementary $y$-dipole one lattice spacing along $\hat{x}$.
Just as the hopping terms for single electric charges can be thought of as extending Wilson lines through the translation of a charged particle, hopping a dipole that consists of a pair of charges extends a pair of Wilson lines attached as shown in Fig. \ref{fig:2wilson_lines}(a).
Since a dipole is built from two oppositely-charged monopoles, the Wilson lines that are extended are oppositely oriented, and the net number of Wilson lines passing through any horizontal or vertical line remains unchanged under the action of such a ring-exchange term. However, as we have discussed above, Wilson lines, even when appropriately coupled to charges, are not gauge invariant by themselves in the presence of gauged 1-form symmetries, and we must introduce an additional gauge field to account for this.

In a more general setting, we want to consider lattice models having conserved dipole moments, implemented using global $U(1)$ 1-form electric symmetries acting along lattice lines (in (2+1)d) of a \emph{dual lattice}. We have found that upon gauging these global 1-form symmetries, individual Wilson loops/lines are no longer gauge invariant unless they are attached to a Wilson surface via a coupling to the 2-form field $\mathcal{B}$.  Our central claim is that coupling the dipoles, i.e., the electric lines, to the background 2-form field $\mathcal{B}$ provides a natural interpretation for the appearance of the (off-diagonal) rank-2 lattice gauge field $A_{xy}$ that couples to the ring-exchange terms in a lattice Hamiltonian:
\begin{equation}
\label{eqn:2form_peierls}
     c^\dagger_{\textbf{r}} c^{\phantom{\dagger}}_{\textbf{r}+\hat{x}} c^\dagger_{\textbf{r}+\hat{x}+\hat{y}} c^{\phantom{\dagger}}_{\textbf{r}+\hat{y}}\to\text{e}^{iA_{xy}}  c^\dagger_{\textbf{r}} c^{\phantom{\dagger}}_{\textbf{r}+\hat{x}} c^\dagger_{\textbf{r}+\hat{x}+\hat{y}} c^{\phantom{\dagger}}_{\textbf{r}+\hat{y}}.
\end{equation}

Let us now provide the explicit construction relating the 2-form gauge field $\mathcal{B}$ and the Peierls' factor coupled to the ring-exchange term. 
Since the operator (\ref{eqn:ring-exchange1}), which contains only fermionic fields, has a two-fold interpretation in terms of dipole hopping, we find that the electromagnetic Peierls' phase for both of these processes would yield different results.
For example, resolving (\ref{eqn:ring-exchange1}) as a pair of electronic $\hat{x}$-hopping operators in the background electromagnetic field $\mathcal{A}$ we have:
\begin{equation}
\begin{split}
    & c^\dagger_{\textbf{r}} c^{\phantom{\dagger}}_{\textbf{r}+\hat{x}} c^\dagger_{\textbf{r}+\hat{x}+\hat{y}} c^{\phantom{\dagger}}_{\textbf{r}+\hat{y}}\\
    &\to\left(\text{e}^{-iA_x(\textbf{r})}c^\dagger_{\textbf{r}}c^{\phantom{\dagger}}_{\textbf{r}+\hat{x}}\right)\left(\text{e}^{iA_x(\textbf{r}+\hat{y})}c^\dagger_{\textbf{r}+\hat{x}+\hat{y}}c^{\phantom{\dagger}}_{\textbf{r}+\hat{y}}\right),
    \end{split}
    \label{eqn:dipole_hop2}
\end{equation}
where $A_x(\textbf{r})\equiv\frac{e}{\hbar}\int_{\textbf{r}}^{\textbf{r}+\hat{x}}\mathcal{A}$.
Clearly, this choice of coupling to the background vector-potential renders the ring-exchange term (\ref{eqn:ring-exchange1}) invariant under the regular electromagnetic (0-form) gauge transformations. 
Equivalently, Eq. \ref{eqn:ring-exchange1} can be resolved as a product of two $\hat{y}$-hopping operators:
\begin{equation}
\begin{split}
    &c^\dagger_{\textbf{r}} c^{\phantom{\dagger}}_{\textbf{r}+\hat{x}} c^\dagger_{\textbf{r}+\hat{x}+\hat{y}} c^{\phantom{\dagger}}_{\textbf{r}+\hat{y}}\\
    &\to-\left(\text{e}^{-iA_y(\textbf{r})}c^\dagger_{\textbf{r}}c^{\phantom{\dagger}}_{\textbf{r}+\hat{y}}\right)\left(\text{e}^{iA_y(\textbf{r}+\hat{x})}c^\dagger_{\textbf{r}+\hat{x}+\hat{y}}c^{\phantom{\dagger}}_{\textbf{r}+\hat{x}}\right),
    \end{split}
    \label{eqn:dipole_hop3}
\end{equation}
which is also gauge-invariant.
In the most general form, the ring-exchange term (\ref{eqn:ring-exchange1}) coupled to the background electromagnetic field takes the following form:
\begin{equation}
    \left(a\text{e}^{i\partial_x A_y}+b\text{e}^{i\partial_y A_x}\right)c^\dagger_{\textbf{r}} c^{\phantom{\dagger}}_{\textbf{r}+\hat{x}} c^\dagger_{\textbf{r}+\hat{x}+\hat{y}} c^{\phantom{\dagger}}_{\textbf{r}+\hat{y}},
    \label{eqn:two_factors}
\end{equation}
where the amplitudes $a+b=1$ to recover the initial ring-exchange term (\ref{eqn:ring-exchange1}) in the absence of electromagnetic fields.

Importantly, as we discussed earlier, this model respects a restricted version of electric 1-form symmetry defined over rows and columns of the dual lattice. This generates two independently conserved 1-form charges:
\begin{equation}
    Q^{(x)}(x_0)=\int_{x=x_0}\star \mathcal{F},\quad Q^{(y)}(y_0)=\int_{y=y_0}\star \mathcal{F},
\end{equation}
where $x_0$ and $y_0$ are coordinates of a column and a row of the dual lattice respectively.\footnote{We explicitly limit ourselves to systems which are locally charge-neutral. If the electric fluxes for any two columns are different, then there must be an excess of electric charge localized between them. So the charges $Q^{(x)}$ for all columns and $Q^{(y)}$ for all rows should be the same under our assumptions. In the ring-exchange model, the electric fluxes between pairs of neighboring rows and columns depend on the initial particle fillings of the rows and columns. Since ring-exchange terms conserve charges for rows and columns, it is enough to specify the fillings of each row and each column to be the same.} These charges represent conserved electric lines (equivalently conserved dipole moments) in the $\hat{x}$ and $\hat{y}$-directions, and they generate two independent transformations for the electromagnetic vector-potential on the lattice:
\begin{equation}
\begin{split}
    A_x(t,\textbf{r})\to A_x(t,\textbf{r}) + f^{(x)}_x(t,x_0,y),\\\ A_y(t,\textbf{r})\to A_y(t,\textbf{r}) + f^{(y)}_y(t,x,y_0),
\end{split}
\end{equation} where $f^{(i)}$ is a 1-form gauge transformation of the lattice field $A$ by the symmetry generated by the charge $Q^{(i)}.$ We note that the components $f^{(x)}_y$ and $f^{(y)}_x$ of the 1-form gauge transformations generated by $Q^{(x)}$ and $Q^{(y)}$ respectively, trivially vanish.

We need to attach additional phase factors to the ring-exchange term to compensate for the 1-form gauge transformation. Let us now determine these additional Peierls' phases. To achieve gauge-invariance with respect to 1-form transformations, we couple our theory to a pair of background lattice 2-form fields $B^{(x)}$ and $B^{(y)},$ one for each conserved charge. These fields are defined in terms of a pair of background continuum 2-form fields $\mathcal{B}^{(x)}$ and $\mathcal{B}^{(y)}$ as 
\begin{equation}
        B^{(x)}=\frac{e}{\hbar}\iint_{p_{xy}}\mathcal{B}^{(x)},\,\, B^{(y)}=\frac{e}{\hbar}\iint_{p_{xy}}\mathcal{B}^{(y)},
        \end{equation} where $p_{xy}$ is the plaquette spanned by the ring-exchange term. Under the 1-form gauge transformation the fields $\mathcal{B}^{(x,y)}$ transform as
\begin{equation}
    \begin{split}
        \mathcal{B}^{(x)}\to \mathcal{B}^{(x)}+d\lambda^{(x)},\\
        \mathcal{B}^{(y)}\to \mathcal{B}^{(y)}+d\lambda^{(y)},
    \end{split}
\end{equation} 
and the lattice versions are given by 
\begin{equation}
    \begin{split}
        B^{(x)}\to B^{(x)}+df^{(x)},\\
        B^{(y)}\to B^{(y)}+df^{(y)},
    \end{split}
\end{equation} where $f^{(i)}=\frac{e}{\hbar}\iint_{p_{xy}}\lambda^{(i)}$, and derivatives acting on the lattice fields are actually nearest neighbor differences. We can combine the transformation of the lattice fields $A$ and $B^{(x,y)}$ to restore gauge invariance:
\begin{equation}
\begin{split}
    &\partial_y A_x-B^{(x)}_{yx}\to \partial_y A_x+\partial_y f^{(x)}_x-B^{(x)}_{yx}-\partial_y f^{(x)}_x,\\
    &\partial_x A_y-B^{(y)}_{xy}\to \partial_x A_y+\partial_x f^{(y)}_y-B^{(y)}_{xy}-\partial_x f^{(y)}_y.
\end{split}
\end{equation}  
Hence, for the two processes described by the ring-exchange term, as shown in Eq. \ref{eqn:two_factors}, we need to add the following phase factors 
\begin{equation}
\begin{split}
    &\text{e}^{i\partial_y A_x}\to\text{e}^{i\partial_y A_x-iB^{(x)}_{yx}},\\ 
    &\text{e}^{i\partial_x A_y}\to\text{e}^{i\partial_x A_y-iB^{(y)}_{xy}}
\end{split}
\end{equation} 
as our 2-form Peierls' factors.

In Refs. \onlinecite{ybh2019,dmmh2019dipole} the rank-2 Peierls' factor introduced for the ring-exchange term was assumed to be unique. 
Therefore, to make the connection between the 2-form gauge field $\mathcal{B}$ and the rank-2 tensor field $A_{xy}$ we should require our 2-form Peierls' factor to be independent of our essentially arbitrary choice of whether to resolve the ring-exchange term as a product of a pair of electron hopping processes along $x$ or along $y$. 
To achieve this it is sufficient to identify $B^{(x)}_{yx}=B^{(y)}_{xy}$. 
Since the lattice fields $B^{(x)}$ and $B^{(y)}$ are given by the spatial integrals, we find the important relationship:  
\begin{equation}
    \iint_S\mathcal{B}^{(x)}=-\iint_{S} \mathcal{B}^{(y)},
    \label{eqn:2form_constraint}
\end{equation} 
for any arbitrary two-dimensional region $S$ of the lattice. This gives us the following relationship between the spatial components of two 2-form fields:
\begin{equation}
    \mathcal{B}^{(x)}_{xy}=-\mathcal{B}^{(y)}_{xy}\equiv \mathcal{B}_{xy}.
    \label{eqn:condition_1}
\end{equation}

While we mathematically motivated this condition it admits an intuitive physical interpretation.
The coupling between the ring-exchange term and the electromagnetic  vector-potential can admit two distinct interpretations depending on whether we choose to resolve it as a pair of $\hat{x}$- or $\hat{y}$- electron hopping operators (or even some linear combination of each). 
In the first case, the pair of Wilson line operators entering (\ref{eqn:dipole_hop2}) plays the role of a translation operator for a segment of $\hat{x}$-electric line in the $\hat{y}$-direction. 
When travelling across a plaquette $p_{xy}$ in the positive $\hat{y}$-direction, this segment of electric line obtains the following phase due to the background 2-form field $\mathcal{B}^{(x)}$:
\begin{equation}
    \Delta\phi^{(x)}=\frac{e}{\hbar}\int_{p_{xy}} \mathcal{B}^{(x)}.
\end{equation}
Similarly, we can choose to resolve the ring-exchange term as translating a segment of $\hat{y}$ electric line in the $\hat{x}$-direction.
Computing the phase obtained during this process due to the background field $\mathcal{B}^{(y)}$ we find:
\begin{equation}
    \Delta\phi^{(y)}=-\frac{e}{\hbar}\int_{p_{xy}} \mathcal{B}^{(y)}.
\end{equation}
Note that the relative minus sign appears here because of the reversed orientation of the pair of vectors involved: one vector that sets the direction of the electric line, and the other one that indicates the direction of motion of the line. The first case is a right-handed orientation while the second is left-handed.
Since we require that both of these processes are implemented by the \emph{same} operator, the overall phase appearing from the coupling to background fields must be defined unambiguously, which prompts us to require
\begin{equation}
    \Delta\phi^{(x)}=\Delta\phi^{(y)}.
    \label{eqn:index_symmetry}
\end{equation}
This can be achieved by imposing the constraint (\ref{eqn:2form_constraint}).

We are now able to provide an alternative interpretation to the rank-2 gauge field $A_{xy}$, as a generalized Peierls' phase arising from a background 2-form field:
\begin{equation}
    A_{xy}=\frac{e}{\hbar}\iint_{p_{xy}}dx dy\ \mathcal{B}_{xy},
    \label{eqn:gen_Peierls}
\end{equation} which is one of the primary conceptual results of our article.
Interpreting $A_{ij}$ as the phase acquired by the $x_i$-dipole moving in the $x_j$-direction, it is evident that Eq. (\ref{eqn:index_symmetry}) is a critical ingredient that enforces the symmetry of the indices of $A_{xy}$ in our construction. We note that generically, $\Delta \phi^{(x)}$ and $\Delta\phi^{(y)}$ may be different and then the tensor $A_{ij}$ could support both symmetric and anti-symmetric components. It is also worth noting that our analysis does not carry over to the diagonal components of $A_{ij},$ as dipole moments moving along/parallel their moment do not sweep out Wilson surfaces that would couple to $\mathcal{B}.$

Finally, we wish to make a connection between the spatio-temporal components of the $\mathcal{B}$ field and the temporal component of the electromagnetic field $\mathcal{A}_t$. 
To do so, let us consider the energetic cost of implementing the process described by the ring-exchange term (\ref{eqn:ring-exchange1}). 
First, consider the energy difference between the two configurations of charges that are connected by the operator (\ref{eqn:ring-exchange1}). 
Introducing the electric potential $\varphi=c\mathcal{A}_t$, we can determine the following change in the energy:
\begin{equation}
\begin{split}
    \Delta\mathcal{E}=e\big(\varphi(\textbf{r}+&\hat{x}+\hat{y})-\varphi(\textbf{r}+\hat{y})\\
    &-\varphi(\textbf{r}+\hat{x})+\varphi(\textbf{r})\big)=eca^2\partial_x\partial_y\mathcal{A}_t.
    \label{eqn:el_pot_charge}
\end{split}
\end{equation}
We note, that for the lattice electric potential field $\mathcal{A}_t$, $\partial_x$ takes the role of the lattice derivative along $\hat{x}$: $\partial_x f(x)\equiv(f(x+a)-f(x))/a$.

We will now repeat this calculation from an alternative perspective. First we can examine the ring-exchange term resolved as a pair of $\hat{x}$ electron hopping operators (\ref{eqn:dipole_hop2}) in the presence of a background $\mathcal{B}$ field. From the structure of the contact term between $\mathcal{B}$ and $\mathcal{F}$ (\ref{eqn:FB_contact}), we see that $\mathcal{B}^{(x)}_{tx}$ plays the role of a potential for  $E_x$ electric lines. Thus,  the energy difference between the two electromagnetic field configurations that are connected by the hopping of $x$-oriented dipoles from Eq. (\ref{eqn:dipole_hop2}) is:
\begin{equation}
    \Delta\mathcal{E}_x=-\frac{e}{\mu_0\varepsilon_0 c}\left(\int_{\textbf{r}+a\hat{y}}^{\textbf{r}+a\hat{x}+a\hat{y}}dx\mathcal{B}^{(x)}_{tx}-\int_{\textbf{r}}^{\textbf{r}+a\hat{x}}dx\mathcal{B}^{(x)}_{tx}\right),
\end{equation} where we have used that the electric field created on the link between sites $\textbf{r}$ and $(\textbf{r}+a\hat{x})$  by the operator $\text{e}^{i\hat{A}_x(\textbf{r})}$ is $\langle \hat{E}_x(\textbf{r})\rangle=-e/\varepsilon_0.$
We can now make a connection between $\mathcal{A}_t$ and the 2-form gauge field $\mathcal{B}^{(x)}$ by requiring that the energy difference derived from the lattice electric potential for charges (\ref{eqn:el_pot_charge}) matches the one we just derived from the electric-line potential: $\Delta\mathcal{E}=\Delta\mathcal{E}_x.$ This relationship produces:
\begin{equation}
\begin{split}
    ec&a^2\partial_x\partial_y\mathcal{A}_t\\
    &=-\frac{e}{\mu_0\varepsilon_0 c}\Big(\int_{\textbf{r}+a\hat{y}}^{\textbf{r}+a\hat{x}+a\hat{y}}dx\mathcal{B}^{(x)}_{tx}-\int_{\textbf{r}}^{\textbf{r}+a\hat{x}}dx\mathcal{B}^{(x)}_{tx}\Big)
    \end{split}
\end{equation}
which gives:
\begin{equation}
    a^2 \partial_x\partial_y \mathcal{A}_t=-\iint_{p_{xy}}dxdy\ \partial_y \mathcal{B}^{(x)}_{tx},
    \label{eqn:A0_gradB}
\end{equation}
where $p_{xy}$ denotes the plaquette between the pair of links in question.
Alternatively, we could have resolved the ring-exchange term as a pair of $\hat{y}$ electron hopping operators, and by matching the corresponding $\Delta\mathcal{E}_y$ with (\ref{eqn:el_pot_charge}), we obtain the following relationship between the $\mathcal{A}_t$ and $\mathcal{B}^{(y)}$ fields:
\begin{equation}
    a^2\partial_x\partial_y \mathcal{A}_t=-\iint_{p_{xy}}dxdy\ \partial_x \mathcal{B}^{(y)}_{ty}.
\end{equation}
Finally, we expect both the calculations of $\Delta\mathcal{E}_x$ and $\Delta\mathcal{E}_y$  to be interchangeable since they are two alternative descriptions of the same physics. 
This gives us the following relationship between spatio-temporal components of the pair of two-form fields:
\begin{equation}
    \iint_{p_{xy}}\partial_y \mathcal{B}^{(x)}_{tx} = \iint_{p_{xy}}\partial_x \mathcal{B}^{(y)}_{ty}.
    \label{eqn:condition_2}
\end{equation}
To reach our final result we note that as the $B^{(x)}$ field, by construction, couples only to $\hat{x}$ electric lines, and $B^{(y)}$  only to $\hat{y}$ electric lines, the pair of components $B^{(x)}_{ty}$ and $B^{(y)}_{tx}$ are redundant since they can take arbitrary values without affecting the physics. We fix this redundancy by setting $B^{(x)}_{ty}=B^{(y)}_{tx}=0$. This last requirement allows us to collect these results into the following form:
\begin{equation}
\begin{split}
    \iint_{p_{xy}}&\left(\partial_x \mathcal{B}^{(x)}_{ty} - \partial_y \mathcal{B}^{(x)}_{tx} \right)=a^2 \partial_x\partial_y \mathcal{A}_t\\
    &=-\iint_{p_{xy}}\left(\partial_x \mathcal{B}^{(y)}_{ty} - \partial_y \mathcal{B}^{(y)}_{tx} \right).
    \label{eqn:A0_to_Bti}
\end{split}
\end{equation} The Eqs. \ref{eqn:gen_Peierls} and \ref{eqn:A0_to_Bti} form the crucial connection between the fields $\mathcal{A}_t,A_{xy}$ and the 2-form field $\mathcal{B}\equiv\mathcal{B}^{(x)}=-\mathcal{B}^{(y)}.$ 
 
While the developments of this section may seem purely formal, we find that they allow for some immediate applications. We will discuss some of these applications in the following two sections, but let us first resolve a question we raised in Section \ref{sec:review_Berry} about the analogy between the Berry phase methods for calculating the polarization and quadrupolarization. The Berry phase calculation for the many-body quadrupole moment was discussed in terms of coupling the ring-exchange model to the uniform Peierls' phase $e^{iA_{xy}},$ and then letting $A_{xy}$ evolve from $0$ to $2\pi/N_x N_y.$ We indicated that there were some conceptual issues when trying to treat this uniform factor as arising from gradients of the electromagnetic vector potential. We can now resolve these inconsistencies by treating $A_{xy}$ as arising from the background 2-form field defined as $\mathcal{B}\equiv\mathcal{B}^{(x)}$ (the field  $\mathcal{B}^{(y)}$ is not independent and determined via the relationships (\ref{eqn:condition_1},\ref{eqn:condition_2})). On a periodic spatial torus the Berry phase is developed as $\mathcal{B}$ evolves from the configuration with $\int_{T^2} \mathcal{B}=0$ to $\int_{T^2} \mathcal{B}=2\pi$, which differ by a large gauge transformation carrying a Chern number of unity (\ref{eqn:2form_lgt}). In terms of the gauge fields that couple to the $x$ and $y$ dipole moments, we can see that we are effectively driving $\int_{T^2}\mathcal{B}^{(x)}$ and $\int_{T^2}\mathcal{B}^{(y)}$ from 0 to $2\pi$ and $-2\pi$ respectively. This amounts to slowly twisting the boundary conditions for large $x$- and $y$-electric lines. There is no issue with choosing $\mathcal{B}$ to be uniform, and hence the final configuration produces a uniform rank-2 field associated with each plaquette $p_{xy}$:
\begin{equation}
    A_{xy}=\frac{e}{\hbar}\iint_{p_{xy}} \mathcal{B} = \frac{2\pi}{N_x N_y},
\end{equation}
which is precisely the change in $A_{xy}$ that was introduced to generate quadrupolar Berry phase\cite{dmmh2019dipole}. 
In summary, the Berry phase (\ref{eqn:qp_berry}) can now be understood as a phase picked up by a wave-function as we adiabatically introduce a \emph{large} gauge transformation of the background gauge field $\mathcal{B}$. 
This construction provides a reasoning for why the rank-2 Berry phase is a sensible physical quantity without sweeping under the rug some inconsistencies that arise when one interprets the rank-2 field in terms of derivatives of rank-1 fields in periodic systems.

\section{Application I: Quadrupolar response}
\label{sec:response}
As our next application we plan to explain the quantization of the quadrupole moment under certain symmetries by considering the topological response of a system to a background 2-form $\mathcal{B}$ field. For simplicity, in this section we always work with field variables that are defined on a closed manifold that is periodic in both space and time. 

In one-dimensional systems, the familiar topological $\theta$-term\cite{Goldstone81,Qi2008} encodes the overall charge polarization of the system through the usual coupling of $p_x$ and the electric field $E_x$:
\begin{equation}
    \label{eq:polar_resp}
    S_{P}=\frac{e\theta}{2\pi}\int d\mathcal{A} \equiv \int dt dx\ p_xE_x,
\end{equation}
where $p_x=\frac{e\theta}{2\pi}$ is the charge polarization. On a closed manifold, and assuming $p_x$ is uniform, this integral computes the Chern class of $\mathcal{A}:$  $\frac{\hbar}{e}\int d\mathcal{A}=2\pi n$.  This reveals the periodic nature of $\theta$: shifting it by $2\pi$ shifts the action by an integer multiple of $2\pi$.  Since this leaves the path integral invariant, we can identify $\theta\equiv\theta+2\pi$.
This identification reflects the periodic nature of polarization, in crystalline systems\cite{kingsmith93}, as it is defined mod 1 (or mod $e$ if we include units of charge). Furthermore, by requiring an invariance of the Lagrangian with respect to spatial inversion symmetry, one immediately finds that the value of $\theta$ can take only two different values: $0$ or $\pi$ mod $2\pi$. 
This reasoning sheds a different light on the quantization of the Berry phase in inversion symmetric systems as we saw in Sec. \ref{sec:review_Berry}.

Using our new interpretation of the quadrupolar Berry phase we will argue that this construction can be extended to derive the quadrupolar response in terms of the two-form field $\mathcal{B}$. In Ref. \onlinecite{ybh2019} it was shown that the quadrupolar response in $(2+1)d$ can be realized via a similar $\theta$-term where the quadrupole moment $q_{xy}$ couples to a rank-2 field:
\begin{equation}
\begin{split}
    S_{Q}&=\frac{e\theta}{2\pi}\int dxdydt \ \left(c\partial_x\partial_y \mathcal{A}_t-\frac{\hbar}{e}\partial_t A_{xy}\right)\\
    &\equiv q_{xy}\int dxdydt\  E_{xy},
\end{split}
    \label{eqn:theta_rank2}
\end{equation}
where $q_{xy}\equiv \tfrac{e\theta }{2\pi},$ and $E_{xy}$ is the rank-2 electric field (which is heuristically a gradient of the ordinary electric field). We note that in our convention $A_{xy}$ is defined as a dimensionless field. Since we treat $A_{xy}$ and $\mathcal{A}_t$ exclusively as lattice fields, it is appropriate to replace the spatial integral of $A_{xy}$ by a sum over plaquettes ${\bf{p}},$ and the integral of $\partial_x\partial_y \mathcal{A}_t$ by a sum over sites $\textbf{r}$ in this formula:
\begin{equation}
    S_{Q}=\frac{e\theta}{2\pi}\int dt\left(\sum_{\textbf{r}}a^2c\partial_x\partial_y\mathcal{A}_t(t,\textbf{r}) -\frac{\hbar}{e}\sum_{\textbf{p}} \ \partial_t A_{xy}(t,\textbf{p})\right).
    \label{eqn:quad_resp_hby}
\end{equation}

All of the work at the end of the previous section will let us translate this response term to the language of 2-forms.
We consider a system with conserved dipole moments in the $x$ and $y$ directions, and we will again take $\mathcal{B}\equiv \mathcal{B}^{(x)},$ and note that the field $\mathcal{B}^{(y)}$ is determined using the relationships (\ref{eqn:condition_1},\ref{eqn:condition_2}). We can make the first replacement: \begin{equation}
   \frac{\hbar}{e}\sum_{\textbf{p}} \partial_t A_{xy}= \sum_{\textbf{p}}\int_{\textbf{p}}dxdy\ \partial_t \mathcal{B}_{xy}=\int dxdy\ \partial_{t}\mathcal{B}_{xy},
    \label{eqn:int_to_sum}
\end{equation} where the admittedly confusing notation $\sum_{\bf p}\int_{\bf p}$ means we are  integrating over the area of each plaquette and then summing them up one at a time. Similarly, using the relationship (\ref{eqn:A0_to_Bti}) between the spatio-temporal components of $\mathcal{B}$ we find:
\begin{equation}
   \sum_{\textbf{r}}a^2\partial_x\partial_y \mathcal{A}_t= \int dxdy \left(\partial_x \mathcal{B}_{ty}-\partial_y \mathcal{B}_{tx}\right).
\end{equation}
Substituting both of these sums back into (\ref{eqn:quad_resp_hby}) we find \begin{eqnarray}
S_Q&=&\frac{e\theta}{2\pi}\int dtdxdy\left(c\partial_x \mathcal{B}_{ty}-c\partial_y \mathcal{B}_{tx}-\partial_t \mathcal{B}_{xy}\right)\nonumber\\&=&-\frac{e\theta}{2\pi}\int d\mathcal{B}.
\label{eqn:theta_2form}
\end{eqnarray} Remarkably, we find that the rank-2 quadrupolar response can be written as a clear topological $\theta$-term response of the 2-form field $\mathcal{B}.$

Thus, with the identification that  $q_{xy}=e \theta/2\pi$ (cf. Eq. \ref{eq:polar_resp}) we propose that the off-diagonal quadrupolar response is manifestly topological when written in terms of $\mathcal{B}$ \begin{equation}
\label{eqn:quad_resp}
    S_Q=-q_{xy}\int d\mathcal{B},\  S_Q \in 2\pi q_{xy}\mathbb{Z}.
\end{equation} This type of topological term computes the  Dixmier-Douady class of $\mathcal{B}$, which is the generalization of the Chern class to $BU(1)$ 2-bundles [\onlinecite{Komargodski19,Murray96,Palumbo2019}]. 
The key result of this formulation is that the integral in Eq. \ref{eqn:quad_resp} is topologically quantized when integrated over a closed 3-manifold, thus demonstrating the topological nature of the quadrupolar response term.

\begin{figure}
    \centering
    \includegraphics[width=0.49\textwidth]{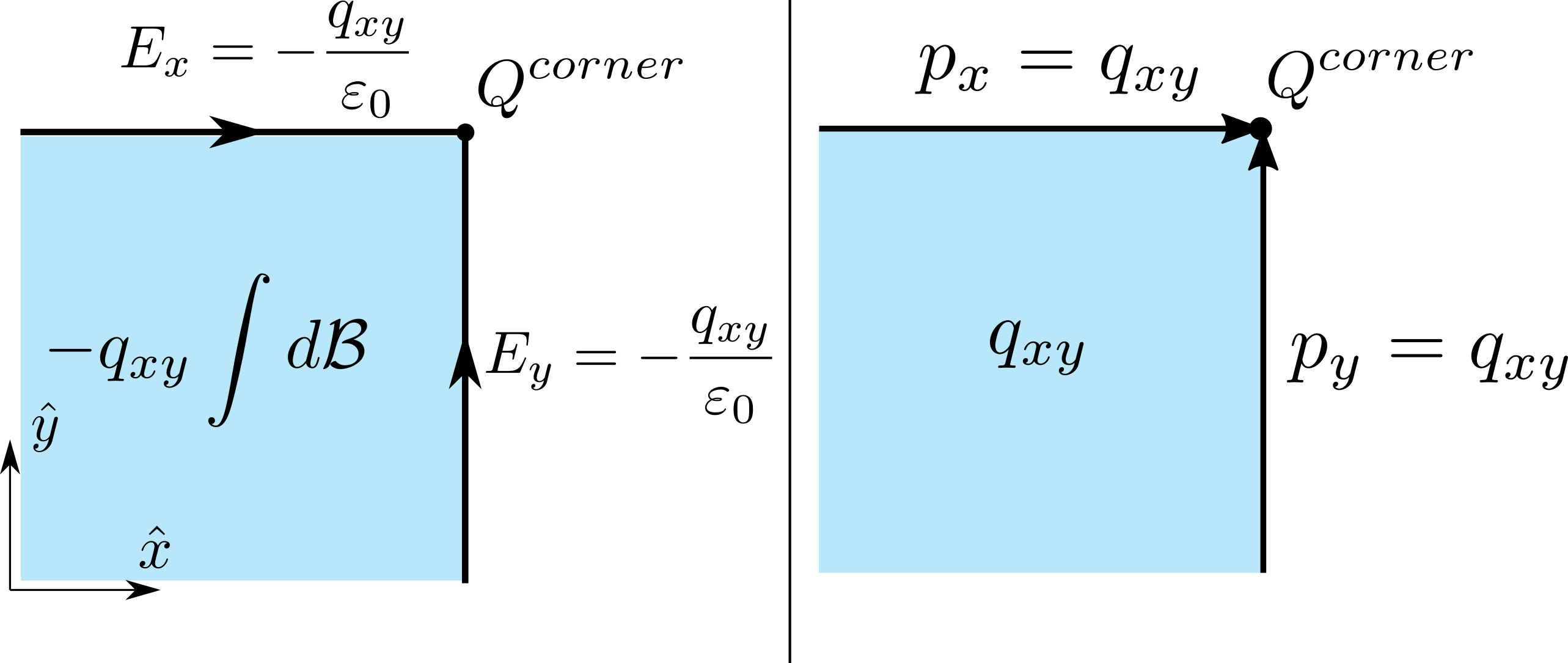}
    \caption{Edge polarizations induced by the bulk Dixmier-Douady term with the coefficient $q_{xy},$ and the edge polarization pattern near the corner of the QTI model\cite{benalcazar2017electric} with the non-zero bulk quadrupolar polarization $q_{xy}$.}
    \label{fig:pol_pattern}
\end{figure}

Now that we have a firm basis for the topological nature of the response action, we can argue that enforcing certain symmetries can quantize the coefficient of the response term.
Since we expect the Maxwell action Eq. \ref{eqn:lagrangian} to be invariant under mirror symmetries and rotations, the individual components of the $\mathcal{B}$ field will inherit their symmetry transformation properties from the corresponding electromagnetic field components through their coupling in Eq. \ref{eqn:lagrangian}.
In particular, consider the mirror symmetry $M_x: x\to -x$.
Under this transformation the electric field component $E_x$ and the magnetic field $B_z$ both invert their signs, while the $E_y$ component of the electric field stays invariant. 
This gives us the following transformation rules for the components of $\mathcal{B}$:
\begin{equation}
    M_x:\ B_{tx}\to -B_{tx},\ B_{ty}\to B_{ty},\ B_{xy}\to -B_{xy}.
\end{equation}
The integrand in Eq. \ref{eqn:quad_resp} then changes its overall sign under the mirror symmetry $M_x$:
\begin{equation}
\begin{split}
    M_x:\ (\partial_t B_{xy}/c&+\partial_y B_{tx}-\partial_x B_{ty})\\
    &\to -(\partial_t B_{xy}/c+\partial_y B_{tx}-\partial_x B_{ty}).
    \end{split}
\end{equation}
This analysis can be similarly repeated for the $M_y:y\to -y$ symmetry with a similar result.
Additionally, the combination $d\mathcal{B}$ is \emph{odd} under $C_4$ rotations. 
This can be seen by noticing that under such rotations we have $\mathcal{B}^{(x)}\to\mathcal{B}^{(y)}$  and the constraints imposed on this pair of fields in the previous section effectively require that:
\begin{equation}
    \left(d\mathcal{B}^{(x)}\right)_{txy}=-\left(d\mathcal{B}^{(y)}\right)_{txy}.
\end{equation}

Now, requiring the action with the quadrupole response term (\ref{eqn:quad_resp}) to be invariant  (modulo $2\pi$) under these symmetries, we obtain a quantization condition on the $\theta$ parameter to take only the values $\theta= n\pi$ for some integer $n$.
This directly translates into a quantized value of the quadrupole moment $q_{xy}$: we find that the distinct allowed values of $\theta$ are $0$ and $\pi$ mod $2\pi$, which translates to a pair of quantized values of the quadrupole moment with $q_{xy}=0$ or $e/2$ modulo $e$. One can also check that the response action (\ref{eqn:quad_resp}) is even under $C_2$ rotations and time-reversal symmetries, which leave the quadrupole moment invariant and would not quantize $\theta, q_{xy}.$
As a result, we see that the $\theta$ coefficient in our response action is quantized by the same symmetries that quantize the quadrupole moment.

Next, let us see what physical phenomena we can derive from this response action. First we can consider the effects of this term in the presence of a boundary which separates the vacuum ($y>0$) from the two-dimensional $(y<0)$ system with a non-vanishing quadrupolar response. At the boundary the response term naturally reduces to: 
\begin{equation}
    S_{\partial}=-q_{xy}\int_{y=0} \mathcal{B}=-\frac{ec\theta}{2\pi}\int_{y=0}dt dx \mathcal{B}_{tx}.
    \label{eqn:Sq_bdy}
\end{equation}
As was noted in the previous section, the $\mathcal{B}_{tx}$ that enters the integral (\ref{eqn:Sq_bdy}) effectively plays the role of a  potential for the electric lines running in the periodic $\hat{x}$-direction. By taking a functional derivative of Eq. \ref{eqn:Sq_bdy} with respect to $\mathcal{B}_{tx},$ and accounting for the coupling constant between the electric field and $\mathcal{B}$ (\ref{eqn:FB_contact}), we find the electric line charge density at the boundary $\rho_{\text{El}}^x=-q_{xy}\mu_0 c^2=-q_{xy}/\varepsilon_0$. As was noted in Section \ref{sec:2form_description}, closed non-contractible electric lines in (1+1)d systems effectively count the polarization, and so we arrive at the quadrupole version of the polarization-surface charge theorem, where we reproduce the connection between a bulk quadrupole moment $q_{xy}$ and a boundary polarization $p_x$:
\begin{equation}
    p_x=-\epsilon_0 \rho_{\text{El}}^x=q_{xy}=\frac{e\theta}{2\pi}.
\end{equation}
Considering the boundary defined by $x=0$ (where the bulk interior is at $x<0$) we find a similar relationship between  $q_{xy}$ and the boundary polarization $p_y$:
\begin{equation}
    p_y=q_{xy}=\frac{e\theta}{2\pi}.
\end{equation} 
To arrive at this result we need to use  Eq. \ref{eqn:condition_2} to turn the integral over $\partial_y\mathcal{B}^{(x)}_{tx}$ into  one over $\partial_x\mathcal{B}^{(y)}_{ty}$, where  $\mathcal{B}^{(y)}_{ty}$ plays the role of a potential for electric $y$-lines.
Note that both $p_x$ and $p_y$ have the same sign relative to the coordinate of their respective one-dimensional manifolds. 

For systems with two edges, one horizontal and one vertical, this creates exactly the pattern of edge polarizations found in the quadrupole topological insulator\cite{benalcazar2017quantized} where the $p_x$ and $p_y$ polarizations `meet' at the corner as shown in Fig. \ref{fig:pol_pattern}. 
The corner charge can be calculated using the relationship (\ref{eqn:A0_to_Bti}) which explicitly introduces the electric potential into the response action, and then following the derivation in Ref. \onlinecite{ybh2019}, where one models the physical corner of the lattice as a product of two step functions, one finds the charge response:
\begin{equation}
    j_0=\frac{e}{2\pi}\partial_x\partial_y\theta(x,y)=\frac{e\theta}{2\pi}\delta(x-x_0)\delta(y-y_0),
\end{equation}
which tells us that there is exactly $\frac{e\theta}{2\pi}$ electric charge localized at the corner with coordinates $(x_0,y_0)$.

Finally, following the recent work [\onlinecite{song19}], we can propose a natural generalization of our quadrupolar response term (\ref{eqn:quad_resp}) to an arbitrary spatial dimension. Similar to the original work that generalized the electric polarization response to higher dimensions, we start by ``gauging the translational symmetry"\cite{thorngren18} by introducing a translation $\mathbb{Z}$-gauge field $\tilde{x}_i$ 
for each $i$-th spatial direction (see Refs. \onlinecite{thorngren18,song19} for details on this notation and the properties of the $\tilde{x}_i$). The quadrupolar response term in $(d+1)$ dimensions then reads:
\begin{equation}\begin{split}
    S_Q=\sum_{i<j}(-1)^{j-i-1}q_{ij}\int d\mathcal{B}\wedge \tilde{x}_1&\wedge ...\tilde{x}_{i-1}\wedge \tilde{x}_{i+1}...\\
    &...\tilde{x}_{j-1}\wedge \tilde{x}_{j+1}...\wedge \tilde{x}_d.
    \end{split}
\end{equation}
We note that the Wilson loop of $\tilde{x}_i$ computed in the $\hat{x}_i$ direction simply gives the length of the lattice in the $i$-th direction: $\int_i \tilde{x}_i=L_i$. Translating this action back from the language of 2-form gauge fields to the rank-2 interpretation, we simply get the following form of rank-2 response action for the quadrupole moment in higher dimensions:
\begin{equation}
    S_Q=\sum_{i<j}\frac{V}{L_i L_j}q_{ij}\int dtdx_idx_j\ E_{ij},
\end{equation}
where $E_{ij}$ is a rank-2 electric field component acting along the $i-j$ plane.

\section{Application II: Higher-form Lieb-Schultz-Mattis Theorem}\label{sec:lsm}

Now that we have identified a method to calculate the electric polarization of dipole conserving systems using electric lines, we are in a position to derive a non-perturbative condition on the ground state degeneracy of dipole-conserving systems, i.e., systems that respect global electric 1-form symmetry. This condition takes the form of Lieb-Schultz-Mattis theorem\cite{lieb1961}, and provides a no-go constraint for the existence of a unique ground state in dipole-conserving systems (as we have defined them in this article).
Similar constraints were recently obtained for systems with $U(1)$ subsystem symmetries\cite{yizhi2019,dmmh2020}, however, as we will see shortly, our newly introduced framework allows us to impose a more generic and stringent condition on the ground state polarization of a dipole-conserving system to ensure the ground state is not degenerate.

\begin{figure}
  \includegraphics[width=0.3\textwidth]{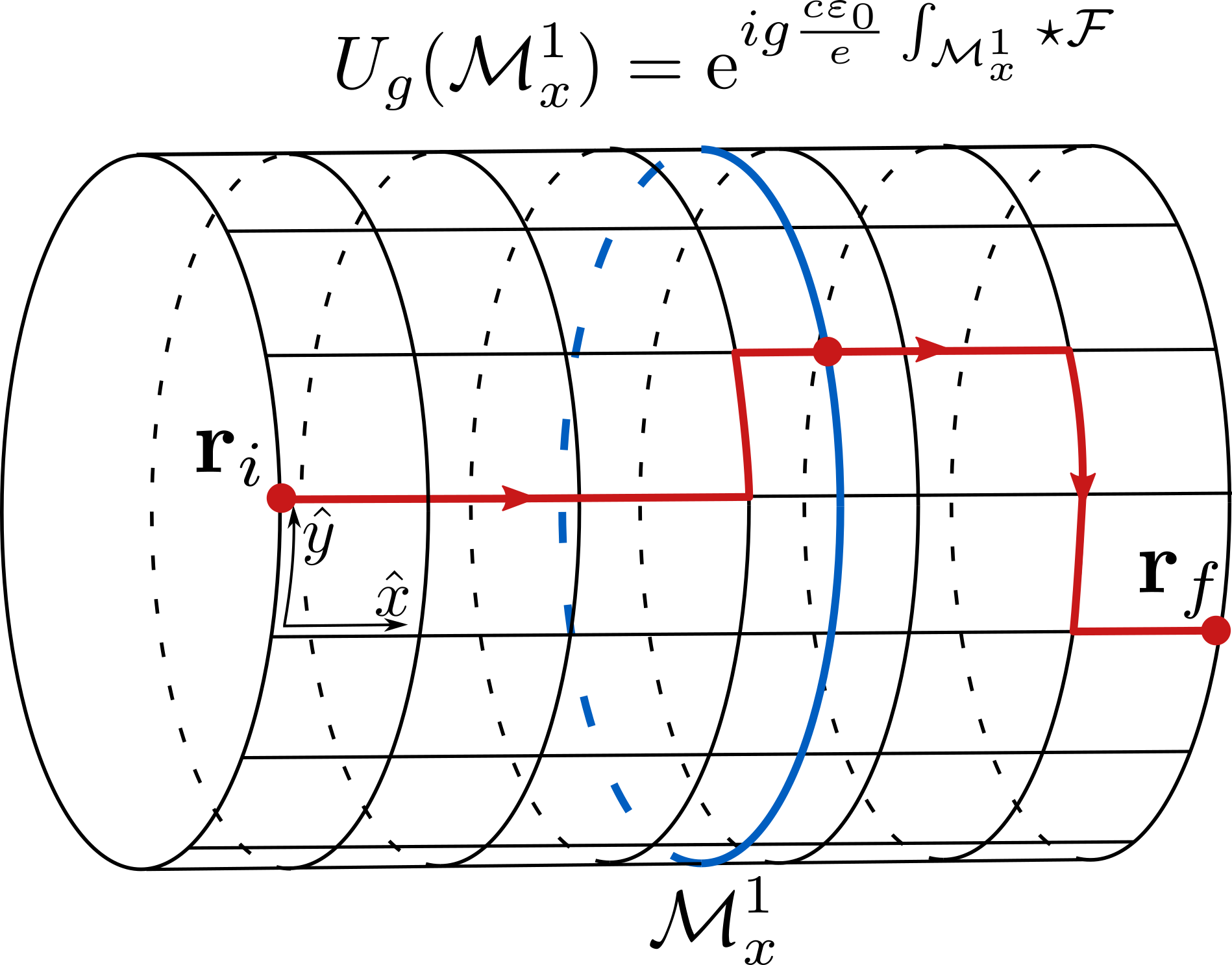}
  \caption{A segment of a periodic square lattice with a 1-form symmetry acting along the blue line $\mathcal{M}^1_x$. For each electric line that passes through $\mathcal{M}^1_x$, the 1-form operator $U_g(\mathcal{M}^1_x)$ assigns a corresponding phase $\text{e}^{ig\Phi_E}$.}
\label{fig:cylinder}
\end{figure}

Before we can proceed with the derivation of the main results of this section, it is worthwhile to revisit the definition of electric polarization introduced in Section \ref{ssec:1form-polarization} in the context of lattice models.
Consider an $N_x\times N_y$ square lattice that is open along $\hat{x}$ and periodic in the $\hat{y}$-direction, hence forming a cylinder with a pair of edges at $x=0$ and $x=L_x$. Let us work with quantized electromagnetic fields on the lattice such that the electric-field number operator $\hat{E}_x(\textbf{r})$ is the canonical conjugate of the lattice field $\hat{A}_x(\textbf{r})$. 
In this context the global electric 1-form symmetry operators are defined over closed one-dimensional manifolds passing over links of the \emph{dual} lattice. Explicitly, let us pick a closed loop $\mathcal{M}^1_x$ on the dual lattice as shown by the blue line in Fig \ref{fig:cylinder}. The 1-form symmetry operator associated with this loop is:
\begin{equation}
    U_g(\mathcal{M}^1_x)=\exp\left(ig\frac{c\varepsilon_0}{e}\int_{\mathcal{M}^1_x}\star \mathcal{F}\right).
    \label{eqn:1form}
\end{equation}
On the lattice this operator takes the following form:
\begin{equation}
    U_g(\mathcal{M}^1_x)=\exp\left(ig\sum_{n=1}^{N_y}\hat{E}_x(\textbf{r}+na\hat{y})\right),
    \label{eqn:1form_elect}
\end{equation}
where we the number operator $\hat{E}_x(\textbf{r})$ counts the amount of electric flux passing along the link connecting neighboring sites with coordinates $\textbf{r}$ and $\textbf{r}+a\hat{x}$.

Now, let us take an insulating state  of electrons $\ket{\Psi(\textbf{r}_i)}$ at a filling factor such that the system is charge neutral (when including ionic charges). Then, imagine a process in which one transports an electron at one of the cylinder's edge sites with the initial position $\textbf{r}_i=(0,y_i),$ to a site $\textbf{r}_f=(L_x,y_f)$ on the opposite edge of the cylinder  along a path $\Xi$ (see Fig. \ref{fig:cylinder}). 
This can be done by applying a sequence of hopping operators which results in the single-electron translation operator that takes the following form:
\begin{equation}
    T_\Xi=c^\dagger_{\textbf{r}_f}\text{e}^{i\sum_{l\in \Xi}\hat{A}_l}c_{\textbf{r}_i}.
    \label{eqn:hopping_Gamma}
\end{equation}
For simplicity, consider the case when $y_i=y_f.$ Since the original state $\ket{\Psi(\textbf{r}_i)}$ was uniformly charge-neutral, the final state $\ket{\Psi(\textbf{r}_f)}\equiv T_\Xi\ket{\Psi(\textbf{r}_i)}$ has a hole at a site $\textbf{r}_i$ and an extra electron at $\textbf{r}_f$. Thus the translation operator $T_\Xi$ can be regarded as an operator that introduces an overall dipole moment $\textbf{d}=-eL_x\hat{x},$ and an $\hat{x}$-polarization equal to $p_x=-\frac{e}{L_y}$ to the system.
As was discussed in Section \ref{ssec:1form-polarization}, the Peierls' phase factor entering $T_\Xi$ plays the role of a creation operator for the quantum electric field line that stretches along $\Xi$.
This electric line can then be detected by calculating the expectation value of the electric flux operator $\hat{\Phi}^x_E=\frac{e}{\varepsilon_0}\sum_{y=1}^{N_y}\hat{E}_x(x_0,y)$ 
for the state $\ket{\Psi(\textbf{r}_f)}$ to obtain an alternative derivation of the expected result: 
\begin{equation}
    p_x=-\frac{L_x}{V}\varepsilon_0\langle\hat{\Phi}^x_E\rangle=-\frac{e}{L_y}.
    \label{eqn:pol_x_ely}
\end{equation}

Let us now introduce periodic boundary conditions in the $\hat{x}$ direction. 
Since we have specified that $y_i= y_f$, the state $\ket{\Psi(\textbf{r}_f)}$ ends up being uniformly charge-neutral after the boundaries are identified. 
As there are no local electric charges in the system, the polarization can no longer be calculated directly from the charge density, however, it can still be evaluated by counting the electric flux created by the translation operator (\ref{eqn:hopping_Gamma}). 
For a closed $\Xi$, $T_\Xi$ acts non-trivially on only the electric-field degrees of freedom living on the links to create an electric line winding around the lattice along the $\hat{x}$-direction. 
Since introducing the periodic boundary conditions does not affect the electric line in the bulk of the system, we expect to obtain the same value of $p_x$ when evaluating it by counting the electric flux as in Eq. \ref{eqn:pol_x_ely}.
However, after transporting an electron around a closed loop we return the electric charge configuration back to the initial one and, therefore, we should not expect any physically observable changes in the overall electric polarization. 
This seeming contradiction is resolved by the ambiguity in Eq. \ref{eqn:Pol-1form}, where electric flux entering the expression for the electric polarization $p_x$ is defined modulo 
$e/\varepsilon_0$ - the amount of electric flux one gains by translating a single electron around the lattice.
Therefore, in the completely generic periodic system we must define the electric polarization with respect to the polarization quantum $e/L_y$ for $p_x$ and $e/L_x$ for $p_y,$ giving us $p_x=-\frac{e}{L_y}=0 \mod \frac{e}{L_y}$.

Now we want to additionally require lattice translation symmetry in both directions.
First, consider the implications of translations $T_x$ acting \emph{along} the direction in which the path $\Xi$ winds around. We find that to satisfy translation symmetry $\Xi$ must map to itself under the action of $T_x$, and therefore, $\Xi$ has to be represented by a straight line running parallel to $\hat{x}$.
Second, to have invariance of the state with respect to lattice translations along $\hat{y}$ -- the direction transverse to $\Xi$ -- we require that there must be an electric line parallel to $\Xi$ in every unit cell in the $\hat{y}$-direction.
These constraints modify the value of polarization quantum that we just proposed\cite{vanderbilt1993}. A state resulting from the process of translating a single electron around a lattice period is now prohibited as it traces a single electric loop which does not respect both lattice translations.
On the other hand, we can consider processes in which $N_y$ electrons are simultaneously translated around the $\hat{x}$-direction, creating $N_y$ closed electric loops running parallel to $\hat{x},$ and spaced out by the lattice translation along $\hat{y}$.
Again, since there is no change in the positions of the collection of electric charges, we must find that the resulting state must have a vanishing polarization up to the polarization quantum.
Calculation of the electric polarization $p_x$ by evaluating the amount of the electric flux winding in the $\hat{x}$-direction  results in a polarization of $p_x=-N_y e/L_y=e/a.$ Thus  we find the necessity of an enlarged polarization quantum for 2D systems respecting translation symmetries of the lattice:
\begin{equation}
    p_x\equiv p_x + \frac{n_x e}{a_y},\,\,\,\, p_y\equiv p_y+\frac{n_y e}{a_x},
    \label{eqn:transl_inv_quantum}
\end{equation}\noindent where $n_x, n_y$ are integers, and we have restored the possibility of anisotropic lattice constants to show the distinct polarization quanta in the two directions.

\subsection{Derivation of the Higher-Form LSM theorem}
Now that we have given a detailed discussion of our lattice definition of the electric polarization, let us proceed to derive a non-perturbative constraint for the polarization of a unique ground state of a dipole-conserving system.
In this subsection we will focus on systems which, on top of the regular global $U(1)$ charge conservation, respect a restricted version of the 1-form electric symmetry, i.e.,  we will require that the Hamiltonian commutes with a set of operators $U_g(\mathcal{M}^1)$, where $\mathcal{M}^1$ are straight lines going along ether one of the rows or one of the columns of the dual lattice; one example of such a symmetry operator is depicted in Fig. \ref{fig:cylinder}. If a different row or column is chosen it will have no effect on the results, as one might expect from translation symmetry.
We note that in the next subsection we provide a generalization for systems that respect the full electric 1-form symmetry defined over arbitrary closed loops instead of this restricted version.

We will consider Hamiltonians that are coupled to quantized electromagnetic degrees of freedom which are represented using bosonic ladder operators (\ref{eqn:bos_ladder}) that act on the links of the original lattice and create/destroy electric lines: 
\begin{equation}
    a_\alpha^\dagger(\textbf{r})=\text{e}^{i\hat{A}_\alpha(\textbf{r})},\ a_\alpha(\textbf{r})=\text{e}^{-i\hat{A}_\alpha(\textbf{r})}.
\end{equation} 
On a square lattice we consider a restricted global 1-form symmetry that acts over a set of lines that run parallel to either the rows or columns of the lattice, and which wind around the torus either in the $\hat{x}$ or $\hat{y}$-directions respectively:
\begin{equation}
    U_{\alpha,g}(\textbf{r})=\exp\left(ig\sum_{n=1}^{N_\beta} \hat{E}_\alpha(\textbf{r}+n\hat{r}_\beta)\right),\ \alpha\neq\beta,
    \label{eqn:1form_global}
\end{equation} where the sum runs over lattice sites in a particular direction, and the electric field operator is a bosonic number operator $\hat{E}_{\alpha}(\textbf{r})=a^\dagger_\alpha(\textbf{r})a_\alpha(\textbf{r})$ which takes integer values.
If we require that the Hamiltonian $H$ commutes with all such 1-form symmetry operators 
\begin{equation}
    U_{\alpha,g} (\textbf{r}) H U^{-1}_{\alpha,g} (\textbf{r}) = H,
    \label{eqn:Ham_1-form_commutes}
\end{equation}
then we find that the simplest allowed local terms in the Hamiltonian take the form $a^\dagger_{x}(\textbf{r})a_x(\textbf{r}+\hat{y})\mathcal{O}_x(\textbf{r})$ and $a^\dagger_{y}(\textbf{r})a_y(\textbf{r}+\hat{x})\mathcal{O}_y(\textbf{r})$, which are a composite of bosonic hopping operators for the electric field lines that act between neighboring links of the lattice, and the fermionic operators $\mathcal{O}_x(\textbf{r})$ and $\mathcal{O}_y(\textbf{r})$ which act on the sites of the lattice. As an example, the ring-exchange model discussed above will contain terms such as $a^\dagger_{x}(\textbf{r})a_x(\textbf{r}+\hat{y})\mathcal{O}_x(\textbf{r})$, where $\mathcal{O}_x(\textbf{r})=c^\dagger_{\textbf{r}}c^{\phantom{\dagger}}_{\textbf{r}+\hat{x}}c^\dagger_{\textbf{r}+\hat{x}+\hat{y}}c^{\phantom{\dagger}}_{\textbf{r}+\hat{y}}$.
The general form of Hamiltonians we consider here reads:
\begin{equation}
    \begin{split}
        H=t\sum_{\textbf{r}}&\left[a^\dagger_{x}(\textbf{r})a_x(\textbf{r}+\hat{y})\mathcal{O}_x(\textbf{r})\right.\\
        &+\left.a^\dagger_{y}(\textbf{r})a_y(\textbf{r}+\hat{x})\mathcal{O}_y(\textbf{r})+h.c.\right],
    \end{split}
    \label{eqn:gen_Ham}
\end{equation} where $t$ is a tunneling coefficient.
Clearly, the ring-exchange model which we discussed in previous sections can be expressed in the form (\ref{eqn:gen_Ham}).
Generalizations of our results to many other electric flux-conserving models are straightforward. 

Now we can follow Refs. \onlinecite{lieb1961,oshikawa2000} to derive an analog of the LSM theorem. First, we consider the following ``dipole twist'' operator:
\begin{equation}
    U^{d}_{X}=\exp\left(\frac{2\pi i }{L_y}\sum_{n=1}^{N_y}na \hat{E}_x(x_0,y=na)\right).
    \label{eqn:1form_gauged}
\end{equation}
Let $\ket{\Psi_0}$ be the ground state of our system. Then:
\begin{equation}
    \begin{split}
        \bra{\Psi_0}&(U^{d}_{X})^{-1}HU^{d}_{X}-H\ket{\Psi_0}=-t\left(\text{e}^{2\pi i/N_y}-1\right)\\
        &\times\sum_{y=1}\left[\langle a^\dagger_x(x_0,y)a_x(x_0,y+a)\mathcal{O}_x(x_0,y)\rangle +h.c.\right].
   \end{split}
\end{equation} Taking the thermodynamic limit $N_y\to\infty,$ and Taylor expanding in powers of $1/N_y$, we see that the $O(1)$ term vanishes provided that the Hamiltonian and the ground state respect parity or time-reversal symmetry. 
Therefore, the energy of the excitation created by the dipole twist operator $U^d_X$ is $O(1/N_y)$, which vanishes in the thermodynamic limit $N_y\to\infty$. 
Hence, the state $\ket{\widetilde{\Psi}_0}=U^d_X\ket{\Psi_0}$ must lie in the ground state subspace.

Now we want to determine if this twisted state is orthogonal to the ground state with which we started. Assuming no spontaneous breaking of the translation symmetry in the $\hat{y}$-direction, the state $\ket{\Psi_0}$ must be an eigenstate of the lattice translation operator $T_y$:
\begin{equation}
    T_y\ket{\Psi_0}=\text{e}^{ik_y}\ket{\Psi_0},
\end{equation}
where $k_y$ is the many-body crystal momentum along $\hat{y}$. 
On the other hand, applying $T_y$ to the ``twisted'' state $\ket{\widetilde{\Psi}_0}$ we find:
\begin{equation}
    T_y\ket{\widetilde{\Psi}_0}=T_yU^d_XT^{-1}_y T_y\ket{\Psi_0}=\text{e}^{ik_y+2\pi i\nu_y}\ket{\widetilde{\Psi}_0},
\end{equation}
where the crystal momentum shift is given by the average electric flux winding along the $\hat{x}$-direction: 
\begin{equation}
    \nu_y=\frac{a}{L_y}\sum_{y=1}^{N_y}\hat{E}_x(x_0,y)\equiv \frac{\varepsilon_0\hat{\Phi}^x_E}{eN_y}.
\end{equation} Whenever $\nu_y\notin\mathbb{Z}$, the twisted state $\ket{\tilde{\Psi}_0}$  has a different crystal momentum from  $\ket{\Psi_0},$ and therefore they must be orthogonal to each other. This would necessitate a ground state degeneracy. 
In contrast, the condition $\nu_y\in\mathbb{Z}$ does not require a degenerate ground state. When this condition is satisfied it requires a set of values of electric flux $\Phi^x_E$, which all, according to Eq. \ref{eqn:transl_inv_quantum}, translate to the polarization $p_x=0$ modulo the quantum $e/a$.
A similar condition can be derived for the $\nu_x$ - the average electric flux winding around in the $\hat{y}$ direction, i.e., we find that the ground state of a periodic dipole-conserving system can be unique only when the system is completely unpolarized up to a polarization quantum
\begin{equation}
    p_x=p_y=0\mod \frac{e}{a}.
    \label{eqn:pol_cond}
\end{equation} In other words, one could heuristically say that the dipole filling factor must be an integer to have a unique ground state.

A similar constraint on the electric polarization in dipole-conserving systems was recently obtained\cite{yizhi2019,dmmh2020}, for $N_x\times N_y$ square lattices with Hamiltonians that commute with a pair of position space twist operators:
\begin{equation}
U_\alpha=\exp\left(\frac{2\pi i}{L_x}\sum_{\textbf{r}}r_\alpha \hat{n}_{\textbf{r}}\right),\ \alpha=x,y,
\end{equation}
whose eigenvalues can also be related to the electronic polarization\cite{resta1998}. In the type of models we consider here, the commutation conditions for $U_x, U_y$ with $H$ are similar to Eq. \ref{eqn:Ham_1-form_commutes}.
The main difference in the approach we take here is that the dipole twist operator used to obtain the LSM condition in Eq. \ref{eqn:pol_cond} is constructed using 1-form symmetry operators built from electric field operators which have less restrictions imposed on them by the periodicity of the system since they are not directly related to a coordinate operator like the $U_x, U_y$ are. 
This difference allowed us to obtain a much more generic constraint than the one discussed in Refs. \onlinecite{yizhi2019,dmmh2020} - our condition (\ref{eqn:pol_cond}) requires that the polarization in a two-dimensional lattice system must vanish modulo a two-dimensional polarization quantum, whereas the earlier work required either that the polarization on each row/column vanished up to a one-dimensional polarization quantum, or that required a restriction of the aspect ratio of the 2D lattice to complete the derivation.

For a system with a boundary our result also yields an interesting implication.
Let our lattice be periodic along $\hat{y}$ and open in $\hat{x}$.
This will result in a cylinder that has two edges having their normal vectors pointing along the open direction $\textbf{n}=\pm\hat{x}$.
Either of the two edges can be considered as a one-dimensional periodic chain with $N_y$ unit cells. 
Because of the surface-charge theorem relating polarization to boundary charge in crystals\cite{vanderbilt1993,Resta2007}, the condition (\ref{eqn:pol_cond}) on the $\hat{x}$-polarization of the system can be re-expressed as a condition on the charge filling factor of the edge. When the bulk has a unique ground state then the boundary charge density induced by the polarized bulk is given by the equation $\sigma^b=\textbf{p}\cdot\textbf{n}=0$ modulo $e/a$, i.e., the number of electrons per unit cell of the edge chain must be an integer. This is interesting because it implies that for the bulk to have a unique ground state the boundary will need to have integer filling. This condition on the boundary is precisely the conventional condition of the one-dimensional LSM theorem\cite{yamanaka97} that states that a translationally invariant system that conserves the total particle number (and parity or time reversal)  cannot have a unique ground state if the fermion number per unit cell is not an integer.
In other words, a dipole-conserving system that satisfies the dipole LSM condition (\ref{eqn:pol_cond}) so that it can be in a bulk dipole-insulating phase having a unique ground state, must also necessarily satisfy the charge LSM condition on the boundary that allows for a charge-insulating phase at the edge. This is natural because if the bulk was a diople metal then the surface would likely be able to carry charge currents from the surface-charge theorem, whereas if the bulk is a dipole insulator, the surface will not exhibit charge currents and it will be insulating itself.

\subsection{Flux-threading argument for higher-form LSM theorem}
\label{app:flux_threading}
Now let us provide a derivation of the LSM constraint using an alternative method. We will apply a generic flux-threading argument for  2-form fields, which is analogous to the flux-threading argument that was used to derive the charge LSM theorem in Ref. \onlinecite{lu2020filling}.
Here we will relax the restriction placed on the 1-form symmetries in the previous subsection, and will consider systems with Hamiltonians that commute with \emph{all} possible electric 1-form symmetries.
The discussion in this subsection mirrors the one presented in Ref. \onlinecite{Ryu2018}, with the primary difference being that we have a different physical interpretation for the line-like objects charged under the 1-form symmetry.

We will study a periodic two-dimensional system that respects both the global charge (0-form) and dipole (1-form) $U(1)$ symmetries. 
First consider the following dipole twist operator: 
\begin{equation}
    U^d_X=\exp\left(\frac{2\pi ic\varepsilon_0}{eL_y}\int_{\mathcal{M}_y}\star\mathcal{F}\ y\right),
    \label{eqn:dipole_twist_app}
\end{equation}
where $\mathcal{M}_y$ is a closed line (e.g., defined by the equation $x=0$) that wraps around the system in the $\hat{y}$-direction.
The length of $\mathcal{M}_y$ is therefore equal to the corresponding dimension of the system $L_y$. 
When applied to the Hamiltonian of the system, this operator performs a large gauge transformation to the background 2-form field $\mathcal{B}$. Second, let us consider a procedure in which 2-form flux is adiabatically inserted in a fashion that does not break  translation symmetry by defining the Hamiltonian:
\begin{equation}
    \mathcal{H}(\Phi,t)=\mathcal{H}\left(\mathcal{B}_{xy}=\frac{c\varepsilon_0\Phi}{e L_x L_y}\frac{t}{T}\right),
\end{equation}
where $\mathcal{B}_{xy}$ is the (only) spatial component of the 2-form field $\mathcal{B},$ and the time period $T$ taken to be large.
The time-evolution operator for this process is
\begin{equation}
    \mathcal{U}_t(\Phi)=\mathcal{T}\exp\left(-i\int_0^T\mathcal{H}(\Phi,t)\right),
    \label{eqn:ad_flux_ins_app}
\end{equation}
where $\mathcal{T}$ denotes the time-ordering. For $\Phi=2\pi$, this operator inserts one quantum of 2-form flux, which can then be removed by the twist operator \ref{eqn:dipole_twist_app}, and so we can define the combined operator:
\begin{equation}
    \mathcal{U}_t^d=(U^d_X)^{-1}\mathcal{U}_t(2\pi),
\end{equation}
which leaves the Hamiltonian of the system invariant.

Importantly, if we examine the action of this operator on the ground state $\ket{\Psi_0}$, we might find that the state $\mathcal{U}_t^d\ket{\Psi_0},$ which must also lie in the ground state subspace, differs from the original ground state.
Consider the action of the lattice translation operator $\hat{T}_y$ on the ground state:
\begin{equation}
    T_y\ket{\Psi_0}=\text{e}^{ik_y}\ket{\Psi_0},
\end{equation}
where $k_y$ is the total crystal momentum of the ground state along $\hat{y}$.
Now let us examine the action of $T_y$ on the state $\mathcal{U}_t^d\ket{\Psi_0}$.
Since the adiabatic flux insertion (\ref{eqn:ad_flux_ins_app}) is done while preserving the translation symmetry, we have:
\begin{equation}
    [T_y,\mathcal{U}_t(2\pi)]=0.
\end{equation}
On the other hand, commuting $T_y$ with the dipole twist operator $U^d_X$ we find:
\begin{equation}
    T_y U^d_X=U^d_X T_y\exp\left(\frac{2\pi ic\varepsilon_0}{eN_y}\int_{\mathcal{M}_y}\star\mathcal{F}\right).
\end{equation}
The integral in the phase factor quantifies the total electric flux/polarization in the $\hat{x}$-direction and so we have:
\begin{equation}
    T_y \mathcal{U}_t^d\ket{\Psi_0}=\exp\left(ik_y + 2\pi i\frac{\varepsilon_0 \hat{\Phi}^x_{E}}{eN_y}\right) \mathcal{U}_t^d\ket{\Psi_0}.
\end{equation}
Therefore, we once again see that, unless $\nu_y\equiv\varepsilon_0\hat{\Phi}^x_{E}/eL_y$ is an integer, the state $\mathcal{U}_t^d\ket{\Psi_0}$ is orthogonal to the original state $\ket{\Psi_0},$ and the ground state cannot be unique. 
An analogous constraint can be derived for the other component of electric polarization $p_x$, considering the dipole twist operator implemented along one of the rows.
Thus, from this derivation we have arrived at the same pair of conditions for electric polarization (\ref{eqn:pol_cond}) which we previously obtained for a dipole-conserving system on a square lattice.

\section{Discussion and Conclusion}
In this article, we have compared the physics of lattice models augmented by $U(1)$ dipole and/or subsystem symmetries to systems having generalized higher-form global symmetries. 
Focusing on $U(1)$ subsystem symmetries that are supported on co-dimension 1 sublattices, and field theories with 1-form symmetries, we have found a remarkable correspondence between these two physical frameworks. 
In particular, we have shown that a 2-form field that arises as a result of gauging the 1-form electric symmetry can be related to the rank-2 tensor gauge field  via a generalized Peierls' substitution.
We have shown that this 2-form field can be used to define a topological response term for the quadrupole moment which, via our mapping,  translates to the recently introduced quadrupolar response term for rank-2 fields\cite{ybh2019}.
The introduced response term takes the form of a Dixmier-Douady topological invariant, which is a natural generalization of the Chern number for 2-form fields, and provides a natural route to prove the symmetry-enforced quantization of the bulk quadrupole moment.
Furthermore, our proposed framework allows us to relate the recently uncovered rank-2 Berry phase\cite{dmmh2019dipole} with the topological properties of the 2-form field. 
Specifically, we have shown that the adiabatic process which is used to define the rank-2 Berry phase, effectively performs a large gauge transformation of the underlying 2-form gauge field. Finally, we used an alternative definition of the electric polarization for periodic systems to prove a generic Lieb-Schultz-Mattis theorem for systems that conserve charge and dipole moment.

One obvious direction for future work is to clarify how our results extend to higher-form symmetries and subsystem symmetries acting on co-dimension $n$ sublattices. 
A natural guess would be that gauging an $n$-form global symmetry introduces a $(n+1)$-form field that can be used to define a general $n$-th order multipole topological response term.
For example, gauging a 2-form symmetry leads to a 3-form field that can then be associated with a volume trapped inside an elementary ``cube-exchange" term that hops an elementary quadrupole. 
In the language of subsystem symmetries, such cube-exchange terms are exactly the ones allowed by $U(1)$ symmetries conserving charge on every one-dimensional subsystem of the cubic lattice. It also would be interesting to also consider systems that respect magnetic 1-form symmetries which we have not addressed at all.

Another important question is the anomalous behaviour of the multipole topological response terms in the presence of the boundary. 
We briefly discussed how the anomaly inflow from the bulk results in a polarization response term with a coefficient that matches the bulk quadrupole moment as expected.
It would be interesting to explore the relationship between the topological higher-form gauge theories and boundary-obstructed topological phases\cite{khalaf19} where the topologically non-trivial bulk phase manifests in a boundary SPT phase.

Our framework also provides hints at physics beyond multipolar insulators.  Higher-rank gauge theories are a popular and interesting topic in their own right, and show up quite naturally in other settings.  It would be illuminating to draw precise connections between the higher-rank and higher-form gauge theories.  Questions about gauge-invariant combinations and the relation to gravity may find answers by using the opposite description.

\section*{Acknowledgements}
We would like to thank I. Danilenko, D. Else, R. Thorngren, and R.G. Leigh for useful discussions. ADR is funded by DMR grant no. 1653769, and thanks the Institute for Condensed Matter Theory at the University of Illinois, where this work was initiated.
OD and TLH
thank the US National Science Foundation under grant
DMR 1351895-CAR, and the MRSEC program under NSF Award Number DMR-1720633 (SuperSEED) for support.
TLH also thanks the National Science
Foundation under Grant No.NSF PHY-1748958(KITP) for support.

\bibliography{References.bib}

\appendix

\section{1-form Symmetries}
\label{app:1formstuff}
In this appendix we provide a more detailed discussion of some properties of electric 1-form symmetries.
We will focus our discussion in $(3+1)d$ dimensions for illustrative purposes, however, it is straightforward to translate it to other spatial dimensions greater than one. 
Electric 1-form symmetries act on one-dimensional extended objects which generically intersect a spatial manifold associated with a symmetry operator at a collection of points. By dimension counting we see that the symmetry operators must be associated with co-dimension 2 spatial manifolds, which in (3+1)d are two-dimensional surfaces.
Specifically, we are interested in electric 1-form symmetries generated by the charge:
\begin{equation}
    Q(\mathcal{M}^2)=c\int_{\mathcal{M}^2}\star \mathcal{F},
    \label{eqn:charge_op}
\end{equation}
which simply counts the net amount of electric flux piercing through the two-dimensional closed surface $\mathcal{M}^2$. 

Let us consider quantizing the gauge theory  by introducing the following commutation relations between the electric field $\hat{E}_{i}(\textbf{r})$ and the gauge field $\hat{A}_{j}(\textbf{r}')$ operators:
\begin{equation}
    [\hat{\mathcal{A}}_{i}(\textbf{r}),\hat{E}_{j}(\textbf{r}')]=-i\frac{\hbar}{\varepsilon_0}\delta_{\alpha\beta}\delta(\textbf{r}-\textbf{r}').
\end{equation}
This allows us to see that the charge operator (\ref{eqn:charge_op}) generates transformations of the electromagnetic vector potential on the surface $\mathcal{M}^2$.
To see this explicitly, we first use Poincare duality to rewrite the integral in (\ref{eqn:charge_op}) as an integral over the whole space:
\begin{equation}
    \int_{\mathcal{M}^2}\star\mathcal{F}=\int\star\mathcal{F}\wedge M,
\end{equation}
where $M$ is a Poincare dual of $\mathcal{M}^2$: $M=\delta(\textbf{r}-\textbf{r}_0)(n_x dx+n_ydy+n_zdz)$, where $\textbf{r}_0\in \mathcal{M}^2,$ and $\vec{n}=(n_x,n_y,n_z)$ is a unit vector normal to the surface $\mathcal{M}^2$.
Now it is straightforward to write down transformations of $\hat{\mathcal{A}}$ generated by the 1-form charge operator:
\begin{equation}
    \text{e}^{-igQ(\mathcal{M}^2)}\hat{\mathcal{A}}\text{e}^{igQ(\mathcal{M}^2)}=\hat{\mathcal{A}}+\frac{g\hbar}{\varepsilon_0}M.
    \label{eqn:gen_1_formA}
\end{equation}
For example, in the case when $\mathcal{M}^2$ is a boundary of some compact region $\mathcal{D}^3$, we find that the corresponding Poincare dual 1-form is exact $M=dD,$ and thus the corresponding 1-form charge generates a regular gauge transformation, according to (\ref{eqn:gen_1_formA}).
In the case when the 1-form $M$ (the Poincare dual to $\mathcal{M}^2$) is closed, but not exact, $Q(\mathcal{M}^2)$ generates a global 1-form symmetry.
Lastly, we can gauge the 1-form symmetry by allowing $g$ to smoothly vary along the surface $\mathcal{M}^2$. This allows us to generate translations of $\hat{\mathcal{A}}$ by an arbitrary smooth 1-form:
\begin{equation}
    \hat{\mathcal{A}}\to\hat{\mathcal{A}}+\lambda.
\end{equation}

\begin{figure}
    \centering
    \includegraphics[width=0.45\textwidth]{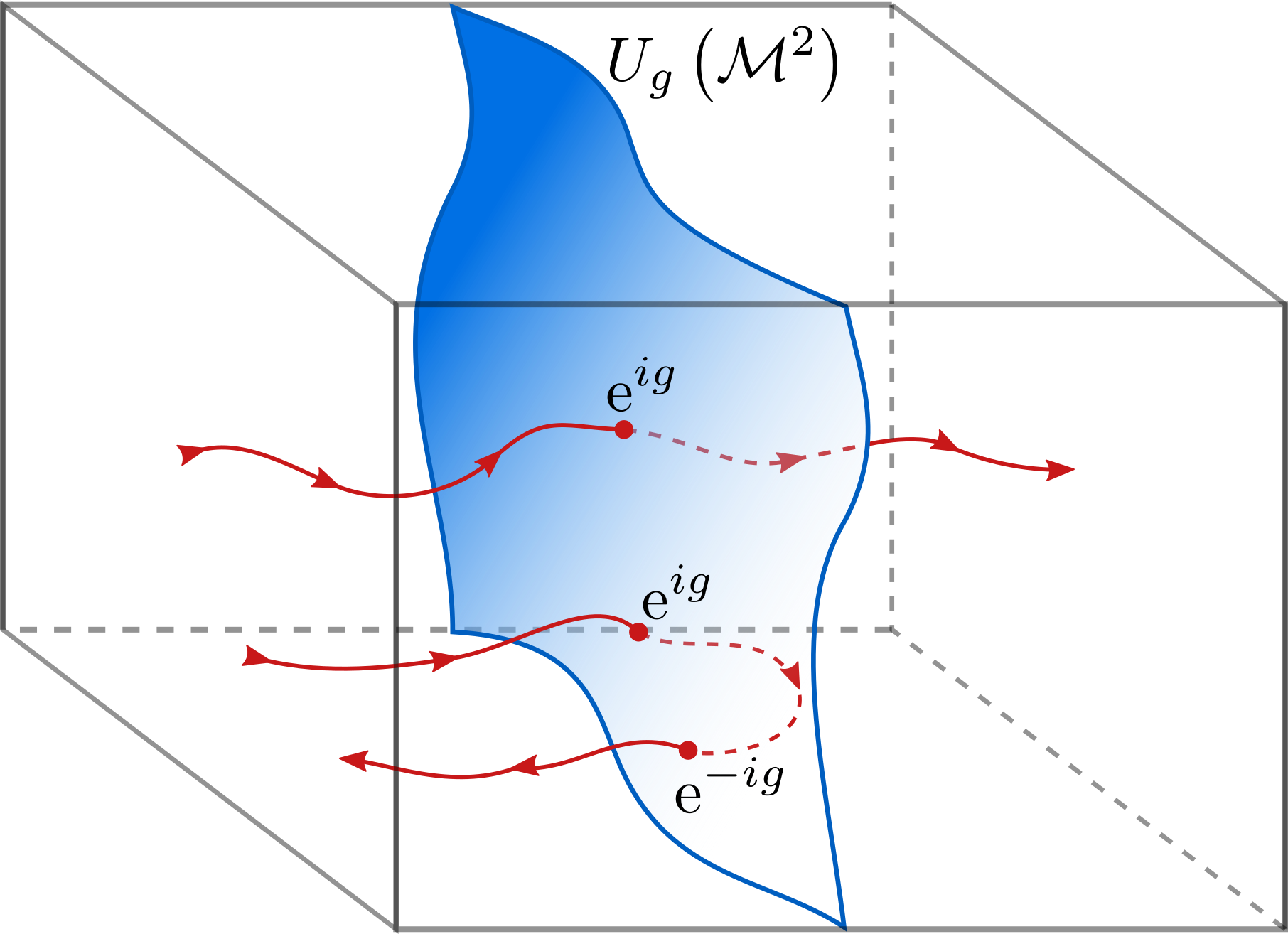}
    \caption{Action of a 1-form global symmetry operator $U_g(\mathcal{M}^2)$ on electric lines created by Wilson line operators. Different orientations of lines relative to the surface $\mathcal{M}^2$ lead to different phase factors: an electric line that pierces the surface twice in two opposite directions contributes no net flux to $\int_{\mathcal{M}^2} \star \mathcal{F}$.}
    \label{fig:surface_wilson_lines}
\end{figure}

Now consider a Wilson loop operator:
\begin{equation}
    \hat{W}_\Gamma=\exp{\left(i\frac{e}{\hbar}\oint_\Gamma \hat{\mathcal{A}}\right)} 
    \label{eqn:wil_loop_op}
\end{equation}
which acts on a state $\ket{0}$ by creating a flux $e/\varepsilon_0$ electric line along a closed path $\Gamma$.
The action of a 1-form symmetry operator $\text{e}^{igQ\left(\mathcal{M}^2\right)}$ can be nontrivial only if the Wilson line goes through the surface $\mathcal{M}^2$. 
The 1-form transformation acts on a Wilson loop operator (\ref{eqn:wil_loop_op}) as:
\begin{equation}
    \text{e}^{-igQ(\mathcal{M}^2)}\hat{W}_\Gamma\text{e}^{igQ(\mathcal{M}^2)}=\text{e}^{ig\Phi_E}\hat{W}_\Gamma,
    \label{eqn:gen_1_form}
\end{equation}
where $\Phi_E$ represents the net amount of electric flux generated by $\hat{W}_\Gamma$ through $\mathcal{M}^2$. This action essentially attaches a phase equal to $g$ times the amount of electric flux generated by $\hat{W}_\Gamma$ through the surface $\mathcal{M}^2$ as shown in Fig. \ref{fig:surface_wilson_lines}. Thus, Wilson lines are charged under the electric 1-form symmetry in exactly the same way that particles are charged under the global charge $U(1)$ symmetry.

In the trivial case, when our $3d$ space is simply-connected (i.e., the fundamental group is trivial), 
any closed Wilson line that passes through a given surface must come back through in the opposite direction. Thus, the net electric flux through $\mathcal{M}^2$ is always zero, and the 1-form symmetry acts on Wilson loop operators as the identity. 
However, in the case when the space supports non-contractible paths (i.e., its fundamental group is non-trivial), this is no longer the case. 
For simplicity, consider a three-torus $T^3$  obtained by taking a cube and gluing its opposing sides. This manifold supports non-contractible paths that wind around one of the three cycles of this space. There also exist closed surfaces $\mathcal{M}^2$ that are sensitive to Wilson lines defined along such non-contractible paths. 
Heuristically, these surfaces ``cut through'' the periodic spatial torus, have no boundary, and do not bound a volume themselves, and thus generate a global 1-form transformation as mentioned above. 
For these surfaces, it is possible that the path $\Gamma$ only passes through $\mathcal{M}^2$ once and reconnects to itself around the periodic direction in space.  
Thus, a 1-form symmetry operator generates non-trivial phases only for such non-contractible Wilson loops.

Interestingly, the non-trivial action of the global 1-form symmetry operator can be intuitively understood as an insertion of a regular magnetic flux through the loop $\Gamma$.
Indeed, consider the following transformation of the electromagnetic field $\mathcal{A}$ by a 1-form $\lambda$:
\begin{equation}
\label{eq:flatshift}
    \mathcal{A} \rightarrow \mathcal{A} + \lambda.
\end{equation}
Examining this transformation in the framework of the regular electromagnetism, we can interpret it as threading an additional amount of magnetic flux equal to $\Delta\phi=\frac{e}{\hbar}\int_\Gamma\lambda$ through the loop $\Gamma$. 

As the state state $\ket{W_\Gamma}$ is obtained by acting with a Wilson line operator (\ref{eqn:wil_loop_op}) on the vacuum, it is evident that for any particular choice of $\lambda$ the Wilson loop operator simply acquires a phase shift equal to $-\Delta\phi$:
\begin{equation}
\begin{split}
    \exp{\left(i\frac{e}{\hbar} \int_\Gamma \mathcal{A}\right)} &\rightarrow \exp{\left(i\frac{e}{\hbar} \int_\Gamma \left(\mathcal{A} + \lambda\right)\right)}\\ 
    =& \exp{\left(i\frac{e}{\hbar} \int_\Gamma \lambda\right)}\exp{\left(i\frac{e}{\hbar} \int_\Gamma \mathcal{A}\right)}.
    \end{split}
\end{equation}
A global 1-form transformation corresponds to a 1-form $\lambda$ such that a Wilson loop operator (\ref{eqn:wil_loop_op}) defined around any path $\Gamma$ winding once in the $\hat{x}$ direction picks up the same phase $\Delta\phi=ge/\varepsilon_0$.
This is equivalent to inserting an additional magnetic flux $\Delta \phi$ through the corresponding hole of our three-torus.

When there are free charges/electrons in the theory, the process of creating an electric flux line by nucleating an electron-hole pair and then dragging them around in the opposite directions is sensitive to the action of 1-form symmetry operators. Specifically, an electron traveling around the contractible loop $\Gamma$ accumulates an Aharonov-Bohm phase
\begin{equation}
    \Delta\phi= \frac{e}{\hbar}\int_\Gamma \mathcal{A}.
\end{equation} 
Since 1-form transformations effectively take $\mathcal{A}$ to $\mathcal{A}+\lambda,$ they shift $\Delta\phi$ by $\frac{e}{\hbar}\int_{\Gamma} \lambda$. 
For an electron to be insensitive to 1-form transformations we must therefore restrict the extra phase shift due to $\lambda$ to satisfy $\int_{\Gamma} \lambda = 2\pi n$. 
If instead we had charge-$p$ matter, there would be additional non-trivial 1-form transformations where $\int_{\Gamma} \lambda = 2\pi n/p,$ meaning that the group of $U(1)$ 1-form transformations gets explicitly broken down to $\mathbb{Z}_p$.
It is clear from here that the existence of charge-1 matter (i.e., free electrons) in the theory completely destroys the electric 1-form symmetry and thus must be prohibited.

\end{document}